\documentclass{aa}
\usepackage[varg]{txfonts}
\usepackage{natbib}
\usepackage{amsmath, amssymb}
\usepackage{graphicx}
\usepackage[breaklinks=true]{hyperref}

\newcommand{\angstrom}{\textup{\AA}}
\DeclareMathOperator\erf{erf}

     \def\aap{A\&A}
     \def\apjl{ApJL}
     \def\apjs{ApJS}

\hypersetup{
	pdftoolbar=false,
	pdfmenubar=true,
	pdfstartview={FitP},
	breaklinks=true,
	frenchlinks=false,
        colorlinks=true,
        linkcolor=red,
        citecolor=blue,
        filecolor=cyan,
        urlcolor=magenta,
        pdftitle={Dynamic mineral clouds on HD 189733b II. Monte Carlo radiative transfer for 3D cloudy exoplanet atmospheres: combining scattering and emission spectra},
	pdfsubject={radiative transfer, planets and satellites: atmospheres, planets and satellites: individual: HD 189733b, methods: numerical},
	pdfauthor={Graham Lee, Kenneth Wood, Ian Dobbs-Dixon, Anna Rice and Christiane Helling}
        }

\begin{document}

\title{Dynamic mineral clouds on HD 189733b}
\subtitle{II. Monte Carlo radiative transfer for 3D cloudy exoplanet atmospheres: combining scattering and emission spectra}
\author{G. Lee\inst{1}\thanks{E-mail:
gl239@st-andrews.ac.uk} \and K. Wood\inst{1} \and I. Dobbs-Dixon\inst{2} \and A. Rice\inst{1}  \and Ch. Helling\inst{1}}
\institute{Centre for Exoplanet Science, SUPA, School of Physics and Astronomy, University of St Andrews, North Haugh, St Andrews, Fife KY16 9SS, UK \and
NYU Abu Dhabi, PO Box 129188, Abu Dhabi, UAE}

\date{Received: 28 Sep 2016 / Accepted: 3 Jan 2017}

\abstract{%Context
As the 3D spatial properties of exoplanet atmospheres are being observed in increasing detail by current and new generations of telescopes, the modelling of the 3D scattering effects of cloud forming atmospheres with inhomogeneous opacity structures becomes increasingly important to interpret observational data. 
}
{%Aims
We model the scattering and emission properties of a simulated cloud forming, inhomogeneous opacity, hot Jupiter atmosphere of HD 189733b.
We compare our results to available Hubble Space Telescope (HST) and Spitzer data and quantify the effects of 3D multiple scattering on observable properties of the atmosphere.
We discuss potential observational properties of HD 189733b for the upcoming Transiting Exoplanet Survey Satellite (TESS) and CHaracterising ExOPlanet Satellite (CHEOPS) missions.
}
{%Methods
We developed a Monte Carlo radiative transfer code and applied it to post-process output of our 3D radiative-hydrodynamic, cloud formation simulation of HD 189733b.
We employed three variance reduction techniques, i.e. next event estimation, survival biasing, and composite emission biasing, to improve signal to noise of the output.
For cloud particle scattering events, we constructed a log-normal area distribution from the 3D cloud formation radiative-hydrodynamic results, which is stochastically sampled in order to model the Rayleigh and Mie scattering behaviour of a mixture of grain sizes.
}
{%Results
Stellar photon packets incident on the eastern dayside hemisphere show predominantly Rayleigh, single-scattering behaviour, while multiple scattering occurs on the western hemisphere.
Combined scattered and thermal emitted light predictions are consistent with published HST and Spitzer secondary transit observations.
Our model predictions are also consistent with geometric albedo constraints from optical wavelength ground-based polarimetry and HST B Band measurements.
We predict an apparent geometric albedo for HD 189733b of 0.205 and 0.229, in the TESS and CHEOPS photometric bands respectively.
}
{%Conclusions
Modelling the 3D geometric scattering effects of clouds on observables of exoplanet atmospheres provides an important contribution to the attempt to determine the cloud properties of these objects.
Comparisons between TESS and CHEOPS photometry may provide qualitative information on the cloud properties of nearby hot Jupiter exoplanets.
}

\keywords{radiative transfer -- planets and satellites: atmospheres -- planets and satellites: individual: HD 189733b -- methods: numerical}

\maketitle

\section{Introduction}
\label{sec:Intro}
%Observational motivation
The 3D spatial properties of exoplanet atmospheres are being observed in increasing rate and detail.
Two key variable for exoplanet characterisation are the properties of their atmospheric cloud complex and how cloud particle size, compositions, global distributions and opacity affect the resultant observations.
A westward offset in the optical phase curves of Kepler-7b was observed by \citet{Demory2013} which was interpreted as cloud particles that backscatter optical photons on the western hemisphere.
But a reduced or absent cloud particle abundance on the eastern hemisphere.
More Kepler planets (e.g.. Kepler-12b, Kepler-8b, Kepler-41b; \citealt{Angerhausen2015,Esteves2015,Shporer2015}) were found to exhibit similar phase curve behaviour, revealing that offset optical phase curves may be a common feature of hot Jupiter atmospheres.
Some Kepler field planets (e.g.. HAT-P-7b, Kepler-5b, Kepler-7b, Kepler-17b, Kepler-41b, Keper-76b; \citealt{Heng2013,Angerhausen2015,Esteves2015}) also have non-negligible geometric albedos (A$_{g}$ $>$ 0.1) in the Kepler bandpass (0.43-0.89 $\mu$m), suggesting the presence of an optical to near-IR scattering aerosol.
B Band (0.29-0.45$\mu$m) and V Band (0.45-0.57$\mu$m) geometric albedo measurements of HD 189733b by \citet{Evans2013} using HST STIS show a blueward slope in the optical, inferring the presence of wavelength-dependent, backscattering cloud particles.

Because of its proximity to its host star, HD 189733b is likely to be tidally locked, with one hemisphere in constant daylight.
Infrared flux phase curve observations of HD 189733b using Spitzer \citep{Knutson2007,Charbonneau2008,Knutson2009,Agol2010,Knutson2012} in photometry bands centred at 3.6$\mu$m, 4.5$\mu$m, 8.0$\mu$m and 24$\mu$m show eastward offsets in peak flux from the sub-stellar point.
Global circulation models and radiative-hydrodynamic models of this hot Jupiter \citep[e.g.][]{Showman2009, Dobbs-Dixon2013, Kataria2016} suggest the development of an equatorial supersonic jet that efficiently advects energy density from the dayside to nightside.
Horizontally propagating gravity waves \citep{Watkins2010,Perez-Becker2013} have also been suggested to contribute to the dayside-nightside energy density advection.
These processes result in an east-west hemisphere inhomogeneity in the temperature profiles for HD 189733b, in line with the Spitzer maximum flux offsets.

These observational and modelling investigations suggest that, for cloud forming hot Jupiters, optical and near-IR wavelength instruments such as HST STIS and Kepler are more sensitive to reflected/scattered light from cloud particles in the atmosphere;
while longer wavelength IR instruments such as HST WF3, NICOMS, and Spitzer are more sensitive to the thermal emission from the planet itself. 
Future optical to near-IR wavelength photometry observational programmes such as TESS (0.6-1.0$\mu$m) and CHEOPS (0.4-1.1$\mu$m), will likely characterise a combination of both scattered light and thermal emission for nearby hot Jupiters.
Examining data from both of these instruments could therefore provide information about the cloud properties of their atmospheres, similar to studies of the Kepler field planets.

%Previous Attempts
In recent years, theoretical and modelling efforts have explored into exploring how the properties of cloud particles in exoplanet atmospheres affect their observed phase curves and geometric/spherical albedos.
\citet{Marley1999} used 1D atmospheric models, combined with a parameterised equilibrium cloud model to predict the albedo spectra of exoplanets.
\citet{Cahoy2010} updated and extended the \citet{Marley1999} approach and presented model scattering albedos for Jupiter and Neptune-like exoplanets at different orbital distances from their host stars.
\citet{Webber2015} employed a similar method to \citet{Cahoy2010} to model a west-east patchy cloud scenario for Kepler-7b to fit the offset optical phase curve \citep{Demory2013}.
\citet{Morley2015} investigated the scattering and thermal emission spectra for photochemical haze and cloud forming Super-Earth atmospheres, comparing their models to the flat transmission spectra of GJ 1214b \citep{Kreidberg2014}.
\citet{Kopparla2016} applied a multiple scattering vector radiative transfer scheme to interpret the polarimetry observations of HD 189733b by \citet{Wiktotowicz2015}.

Recently, studies using 3D global circulation models [GCMs] have explored modelling the atmospheric reflected light phase curves of Kepler field objects.
\citet{Oreshenko2016} performed a GCM experiment using the orbital parameters of Kepler-7b, and applied an equilibrium cloud scheme to model the optical and infrared phase curves offsets of the planet.
\citet{Parmentier2016} performed several GCM model experiments over a grid of equilibrium temperatures (T$_{\rm eq}$ = 1000$-$2200 K), corresponding to the Kepler hot-Jupiter planet parameter space.
They post-processed their model output and applied an equilibrium cloud scheme with a two-stream radiative transfer method to model the reflectivity of clouds across the Kepler bandpass.

Modelling the effects of cloud particles on observable properties of exoplanets using Monte Carlo radiative transfer [MCRT] techniques began with \citet{Seager2000b}, which mainly focused on producing phase and polarised light curves from a model atmosphere.
\citet{Hood2008} applied a parameterised cloud model to a 3D idealised atmosphere of HD 209458b and modelled scattered light curves and geometric albedos of varying cloud heights and scattering properties.
\citet{deKok2012} used a 3D Monte Carlo scattering code to investigate the influence of forward scattering particles on transit spectroscopy.
\citet{Munoz2015} used a Pre-conditioned Backward Monte Carlo method to constrain the cloud particle scattering properties of Kepler-7b.
Recently, Monte Carlo transport methods have also been used to model the path and decay of cosmic rays through a Jupiter-like atmosphere \citep{Helling2016b}, important to the ion chemistry in exoplanet atmospheres \citep{Rimmer2016}.

In this study, we present a 3D Monte Carlo radiative transfer simulation for use in post-processing GCM/RHD atmospheric results with a strong scattering component such as clouds.
Our paper outline is as follows;
Section \ref{sec:Approach} explains our MCRT model framework and our approach to modelling the HD 189733b atmosphere using MCRT.
Section \ref{sec:Method} details our model set-up, input quantities and equations for relating the MCRT output to observable quantities.
In Sect. \ref{sec:Results} we present our model results for scattered and emitted light as well as combining the two components.
We also produce Kepler, TESS, and CHEOPS photometric band predictions for HD 189733b based on our simulation output.
Section \ref{sec:Discussion} contains the discussion and Sect. \ref{sec:Conclusions} summarises key results and conclusions. 

\section{Approach}
\label{sec:Approach}

We present a MCRT code based on the \citet{Hood2008} model framework, with significant additions and alterations to the scheme to capture a more realistic 3D hot Jupiter atmosphere.
In \citet{Lee2016} (Paper I in this series) we simulated the atmosphere of HD 189733b using a 3D radiative-hydrodynamic [RHD] model coupled with a self-consistent cloud formation module.
In the present paper, we post-process our cloud forming RHD results using the MCRT model to compare our simulation to available observational data and make observational predictions for the TESS and CHEOPS photometric bands.

\subsection{MCRT modelling framework}

Monte Carlo radiative transfer is a microphysical, stochastic approach to solving radiative transport problems. 
MCRT simulates the random walk of a luminosity packet (henceforth "L-packet" or "packet") through a medium and describes the interactions (scattering, absorption) of the packet with the surrounding medium by probabilistic sampling.
In MCRT, each initialised packet carries a luminosity fraction $\epsilon_{0}$ / $\Delta t$ [erg s$^{-1}$ ], proportional to a source (e.g.. parent star or planetary atmosphere) of luminosity L$_{\rm source}$ [erg s$^{-1}$].
If N L-packets are initialised by the source of luminosity, the luminosity carried per packet is then \citep[e.g.][]{Lucy1999}
\begin{equation}
\frac{\epsilon_{0}}{\Delta t} = \frac{L_{\rm source}}{N}, 
\label{eq:photon_energy}
\end{equation}
where $\Delta$t [s] is the time over which the MCRT experiment is performed; usually $\Delta$t  = 1 s for time independent MCRT such as the current study.
By tracking the proportion of the luminosity of each L-packet that escapes during the simulation in a certain direction, the total luminosity escaping towards a particular observation direction can be found by summing up the contribution of each N L-packet to the total escaping luminosity.
If the total luminosity budget in the simulation is normalised to the source luminosity (i.e. L$_{\rm source}$ = 1), the total contribution of all packets is then the fraction (i.e. f$_{\rm tot}$ $\in$ (0, 1)) of source luminosity that escaped the simulation boundaries.
To retrieve the luminosity that escaped, the source luminosity is calculated and multiplied by the fraction of luminosity that escaped carried by the L-packets.
This is different from numerically solving the radiative transfer equation via, for example, the two-stream approximation \citep[e.g.][]{Toon1986, Marley1999b, Fortney2006} or ray-tracing method \citep[e.g.][]{Rijkhorst2006}.
Since MCRT tracks the random walk of a packet and its interactions, radiative transfer through complicated 3D geometries, inhomogeneous opacities, and highly multiple scattering regions can be modelled.
MCRT methods have been extensively used to investigate radiative transfer in dusty media, for example, protoplanetary disks \citep[e.g.][]{Whitney2003, Harries2004, Pinte2006, Min2009}, the ISM \citep[e.g.][]{Robitaille2011} and dusty galactic scale problems \citep[e.g.][]{Wood2000}.
MCRT has also been applied to the study and retrieval of Earth based cloud properties \citep[e.g.][]{Mayer2009,Stap2016}.
\citet{Whitney2011} and \citet{Steinacker2013} provide a review of MCRT methods and models used for astrophysical-based problems and \citet{Mayer2009} gives details on Earth based applications.
The heavily irradiated, inhomogeneous opacity, and thermal structures of cloud forming hot Jupiter atmospheres is an ideal application for a MCRT approach.

The beginning of the MCRT L-packet tracking process starts by stochastically assigning initial starting coordinates and direction of a given packet, depending on the location of the luminosity source that is emitting the packet.
For example, a stellar L-packet is initialised randomly at the top of the atmosphere on the dayside of the HD 189733b computational grid in our model.
The wavelength-dependent luminosity carried by these stellar packets is the total monochromatic luminosity incident on the atmosphere, divided by the number of simulated packets ($\epsilon_{\rm 0, inc}$ / $\Delta t$ = L$_{\rm inc, \lambda}$ / N$_{\rm inc}$).
The L-packets emitted from the atmosphere itself are assigned a random starting position within a computational cell volume, where the luminosity is proportional to the total monochromatic luminosity of the atmosphere itself ($\epsilon_{\rm 0, atm}$ / $\Delta t$ = L$_{\rm atm, \lambda}$ / N$_{\rm atm}$).
Initialised stellar packets are assumed to travel in a plane-parallel direction towards the planet, while atmospheric, thermally emitted packets are given an isotropic initial direction. 
The random walk of the L-packet is then determined by stochastically sampling probability distributions that govern the behaviour of the packet (distance travelled, scattering directions etc.). 

From the Beer-Lambert law in radiative transfer theory, the probability of a photon packet passing, without interaction, through a medium of total optical depth $\tau_{\lambda}$ is given by
\begin{equation}
 P(\tau_{\lambda}) = \rm{e}^{-\tau_{\lambda}} ,
\end{equation}
where the optical depth $\tau_{\lambda}$ is defined as 
\begin{equation}
 \tau_{\lambda} = \int^{l_{\rm max}}_{0}\rho_{\rm gas}(l)\kappa_{\rm ext, total}(\lambda, l)\textrm{d}l,
 \label{eq:tau}
\end{equation}
where $\rho_{\rm gas}$ [g cm$^{-3}$] is the local gas density, $\kappa_{\rm ext, total}$ [cm$^{2}$ g$^{-1}$] (Eq. \ref{eq:ktot}) the local total extinction opacity, including absorption and scattering, and $l$ [cm] the path length.
The probability of a packet interacting with the surrounding medium is then
\begin{equation}
 P(\tau_{\lambda}) = 1 - \rm{e}^{-\tau_{\lambda}}.
\end{equation}
The MCRT method stochastically samples this probability distribution with the use of a (pseudo) random number, P($\tau_{\lambda}$) =  $\zeta$ $\in$ [0,1]. 
An optical depth, determined stochastically, encountered by the packet before an interaction is given by
\begin{equation}
\label{eq:tausamp}
 \tau_{\lambda} = -\ln(1 - \zeta).
\end{equation}
Using the definition of $\tau_{\lambda}$ (Eq. \ref{eq:tau}), the distance that the packet travelled through the simulation can then be computed should the density and extinction opacity in the path of the packet be known.
After the packet has travelled the distance given by the sampled $\tau_{\lambda}$, an interaction with the surrounding medium occurs.

\subsection{L-packet interactions with dust and gas}

Once the coordinates specified by the optical depth sampling and direction of travel have been reached by the packet, a scattering or absorption event is determined stochastically.
In cloud forming hot Jupiter atmospheres, the packet interacts with two components: the gas and cloud particles.
An interacting packet exhibits different scattering and absorption behaviours depending what component it is interacting with.
The local probability of the packet interacting with the gas phase $P_{\rm gas}$ is given by
\begin{equation}
P_{\rm gas} = \frac{\kappa_{\rm ext, gas}}{\kappa_{\rm ext, cloud} + \kappa_{\rm ext, gas}},
\end{equation}
where $\kappa_{\rm ext, gas}$ [cm$^{2}$ g$^{-1}$] is the total opacity of the gas and $\kappa_{\rm ext, cloud}$  [cm$^{2}$ g$^{-1}$] (defined by Eq. \ref{eq:cloud_opc}) the total opacity of the cloud component.
Should $\zeta_{1}$  $<$ P$_{\rm gas}$, (where the indexed random number (e.g.. $\zeta_{1}$,$\zeta_{2}$) denotes a sequence of random, independent, numbers for a particular scheme) the packet is assumed to interact with the gas phase.
The type of interaction, scattering or absorption is determined by sampling the single scattering albedo $\omega_{\rm gas}$ of the gas phase, which is the probability of the packet undergoing a scattering interaction.
Since we only consider a H$_{2}$ scattering gas component n the MCRT scheme (Sect. \ref{sec:Method}), the local single scattering albedo of the gas $\omega_{\rm gas}$ is given by
\begin{equation}
\label{eq:gas_alb}
\omega_{\rm gas} = \frac{\kappa_{\rm sca, H_{2}}}{\kappa_{\rm gas,ext}} ,
\end{equation}
where $\kappa_{\rm sca, H_{2}}$ [cm$^{2}$ g$^{-1}$] is the scattering opacity of H$_{2}$ and $\kappa_{\rm gas,ext}$ = $\kappa_{\rm gas,abs}$ + $\kappa_{\rm sca, H_{2}}$ [cm$^{2}$ g$^{-1}$] the total extinction for the gas component. 
Should  $\zeta_{2}$  $<$ $\omega_{\rm gas}$ the packet is Rayleigh scattered by H$_{2}$, otherwise it is absorbed by the gas phase.

If the interaction is with the cloud component (i.e.  $\zeta_{1}$ $>$ $P_{\rm gas}$), the type of interaction is given by sampling the local cloud particle single scattering albedo $\omega_{\rm cloud}$ given by
\begin{equation}
\label{eq:cloud_alb}
\omega_{\rm cloud} = \frac{\kappa_{\rm scat, cloud}}{\kappa_{\rm ext, cloud}},
\end{equation}
where $\kappa_{\rm scat, cloud}$ [cm$^{2}$ g$^{-1}$] is the scattering opacity of the cloud particles and $\kappa_{\rm ext, cloud}$  [cm$^{2}$ g$^{-1}$] the cloud particle total extinction.
Should $\zeta_{3}$ $<$ $\omega_{\rm cloud}$ the packet is assumed to be scattered by the cloud component towards a new direction, otherwise it is absorbed by the cloud particles.

After each scattering event, the next interaction location is determined by Eq. \eqref{eq:tausamp} and the sampling process is repeated.
It is important to note that packets can undergo multiple scattering interactions before getting absorbed, especially in high ($\omega_{\rm gas/cloud}$ $>$ 0.9) single scattering albedo regions.

\subsection{Scattering of L-packets by dust and gas}

During a scattering event, a new direction of travel is given to the packet, which is dependent on the geometric and compositional properties of the interacting material as well as the wavelength of the interacting packet.
A key quantity to consider when modelling a scattering event is the size parameter of the scattering particle given by $x = 2\pi a /\lambda$, where $a$ [cm] is the particle size and $\lambda$ [cm] the wavelength of the L-packet.
The scattering behaviour of an interacting packet can be split into two categories:
\begin{enumerate}
\item x $\ll$ 1 - Rayleigh scattering
\item x $\gtrsim$ 1 - Mie scattering
\end{enumerate}
For Raleigh scattering, the particle size is magnitudes smaller than the wavelength of light interacting with it, resulting in the well-known double lobed scattering phase function where the probability of scattering in the forward and backward directions are equal.
In the Mie regime, the particle size and wavelength of light are of similar magnitudes; the light is typically more forward scattering with a small backscattered component at optical wavelengths.
For a scattered packet in MCRT, the new travel direction is generated stochastically from the scattering properties of the interacting material.
This is determined by sampling the (normalised) scattering phase function $\Phi_{\rm scat}(\theta, \phi)$ of the interacting scattering particle, which is the probability of scattering towards angle ($\theta, \phi$).

For small cloud particles with small size parameters x $\ll$ 1 and H$_{2}$ gas particle scattering, we apply the Rayleigh scattering phase function given by
\begin{equation}
\Phi_{\rm RS}(\theta, \phi) = \frac{3}{16\pi}(1 + \cos^{2}\theta).
\end{equation}
The $\theta$ scattering direction is sampled randomly from this distribution, where a rejection method is applied following the advice in \citet{Whitney2011}.
The $\phi$ scattering angle angle is given by
\begin{equation}
\label{eq:phi}
\phi = 2\pi\zeta,
\end{equation}
which is a uniform sampling of the $\phi$ direction across the 2$\pi$ radian circular circumference.
Rayleigh scattering events are wavelength independent and assumed to be elastic.

When a packet scatters off a cloud particle with size parameter x $\gtrsim$ 1, we apply a Henyey-Greenstein [HG] scattering phase function \citep{Henyey1941}.
The HG phase function is an analytic approximation to the Mie scattering phase function, given by
\begin{equation}
\label{eq:HG}
 \Phi_{\rm HG}(g_{\lambda}, \theta, \phi) = \frac{1}{4\pi}\frac{1 - g_{\lambda}^{2}}{[1 + g_{\lambda}^{2} - 2g_{\lambda}\cos\theta]^{3/2}},
\end{equation}
where $g_{\lambda}$ is the wavelength-dependent scattering asymmetry parameter in the range $-$1 to 1, given by the results of the Mie theory.
This is defined as the mean scattering cosine angle from the relation
\begin{equation}
\label{eq:mean_cosine}
g_{\lambda} = \langle \cos\theta\rangle = \int_{\Omega} \Phi_{\rm HG}(g_{\lambda}, \theta)\cos\theta d\Omega .
\end{equation}
A value of g $<$ 0 indicates a preference for packet backscattering, g = 0 an equal backward/forward scattering and g $>$ 0 forward scattering.
The scattering angle $\cos\theta$ is sampled stochastically from this distribution using the following form
\begin{equation}
\label{eq:HG_scat}
\cos\theta = \frac{1}{2g_{\lambda}}\left[1 + g_{\lambda}^{2} - \left(\frac{1 - g_{\lambda}^{2}}{1 - g_{\lambda} + 2g_{\lambda}\zeta}\right)^{2}\right].
\end{equation}
The sampled $\phi$ direction is given by Eq. \eqref{eq:phi}.
We assume elastic scattering for Mie scattering events.
The HG probability distribution has been shown to be a reasonable approximation to Mie scattering within the optical and near-IR wavelength regime \citep{Draine2003}.
However, a small but not insignificant probability of backscattering is not completely captured by this approximation at optical wavelengths \citep[e.g.][]{Kattawar1975, Draine2003, Hood2008, Dyudina2016}.
To address this, we apply a Two-term Henyey-Greenstein [TTHG] function \citep[e.g..][]{Pfeiffer2008, Cahoy2010, Dyudina2016} given by
\begin{equation}
\Phi_{\rm TTHG}(g_{a}, g_{b}, \theta) = \alpha \Phi_{\rm HG}(g_{a}, \theta) + \beta \Phi_{\rm HG}(g_{b}, \theta), 
\end{equation}
where $\alpha$ is the mainly forward scattering component with g$_{a}$ $>$ 0,  and $\beta$ a mainly backscattering component with g$_{b}$ $<$ 0.
For all parameters, the relation $\alpha$ + $\beta$ = 1 has to be satisfied.
We apply the parameters in \citet{Cahoy2010} where g$_{a}$ = g$_{\lambda}$, g$_{b}$ = -g$_{\lambda}$/2 and $\alpha$ = 1 - g$_{b}^{2}$, $\beta$ = g$_{b}^{2}$.
The TTHG form is commonly used in various atmospheric radiative transfer scattering codes (e.g.. \citealt{Marley1999,Cahoy2010,Barstow2014,Munoz2015,Dyudina2016}) and more qualitatively captures the Mie backscattering lobe at optical wavelengths.

To determine the scattering angle of a TTHG event, the probability of forward scattering is simply given by $\zeta$ $<$ $\alpha$ \citep{Pfeiffer2008} after which Eq. \eqref{eq:HG_scat} with g$_{\lambda}$ = g$_{a}$ can be used, otherwise the scattering is a backscattering event in which g$_{\lambda}$ = g$_{b}$ is applied. 
Although $\alpha$ and $\beta$ are used to determine a forward or backward scattering event, due to the probabilistic sampling of the HG function scattering angle, sampling the $\alpha$ term does not necessarily always result in a forward scattering event, nor $\beta$ a backward scattered event.

Since there is a distribution of cloud particle sizes in each cell, the size of an individual cloud particle interacting with the packet must be stochastically determined from the size distribution properties.
This is required as the size distribution may contain a combination of cloud particles in the Rayleigh and Mie size parameter regimes, resulting in a mixture of packet scattering behaviour.
To capture this scattering behaviour, we perform a stochastic sampling from the cumulative distribution function [CDF] of the size distribution, which is the fractional contribution of each particle size to the total area of the particle ensemble.
The area distribution is used, rather than the size distribution, as this better captures the contribution of each cloud particle radius to the scattering cross-sectional opacity of the size distribution (since $\kappa_{\rm sca} \propto f(a)a^{2}$, Eq. \ref{eq:cloud_opc}).
The cloud particle mean grain size, $\langle a\rangle$ [cm], and mean grain area, $\langle A\rangle$ [cm$^{2}$], of the cloud particles in each RHD cell can be found from the dust moment values, $L_{j}$ \citep{Woitke2003, Woitke2004, Lee2015b}.
Assuming an arithmetic log-normal distribution of cloud particle sizes (e.g. \citealt{Ackerman2001}), the mean $\mu$ and standard deviation $\sigma$ (more specifically their natural logarithms) of the ensemble of particles in each cell is then calculated \citep[e.g.][]{Stark2015}.
The surface area log-normal distribution is given by \citep[e.g.][]{Heintzenberg1994}
\begin{equation}
A(a) = \frac{A_{0}}{a\sigma_{A}\sqrt{2\pi}}\exp\left[-\frac{(\ln(a) - \mu_{A})^{2}}{2\sigma_{A}^{2}}\right]
\end{equation}
which is the log-normal size distribution multiplied by the surface area $4\pi a^{2}$, with the total area $A_{0} = 4\pi n_{\rm d}\exp(2\mu + 2\sigma^{2})$ \citep[e.g.][]{Zender2015}, where $a$ [cm] is the sampled grain size, $n_{d}$ [cm$^{-3}$] the total cloud particle number density, $\mu_{A}$ the arithmetic mean of the area distribution and $\sigma_{A}$ the standard deviation of the area distribution.
The relation between the mean and standard deviation of the size distribution ($\mu, \sigma$) and area distribution ($\mu_{A}, \sigma_{A}$) are given by  \citep[e.g.][]{Heintzenberg1994}
\begin{eqnarray}
\mu_{A} = \mu + 2\sigma^{2} , \nonumber \\
\sigma_{A} = \sigma.
\end{eqnarray}
The cloud particle area cumulative distribution function of the log-normal size distribution $A_{\rm CDF}$ $\in$ (0,1) is then constructed in each cell from \citep[e.g.][]{Zender2015}
\begin{equation}
A_{\rm CDF}(a) = \frac{1}{2} + \frac{1}{2}\erf\left[\frac{\ln (a) - \mu_{A}}{\sqrt{2}\sigma_{A}}\right], 
\end{equation}
where $\erf$ is the error function.
We construct $A_{\rm CDF}$ for 100 log-spaced cloud particle size $a$ bins between a minimum seed particle size $a_{\rm seed}$ $\sim$ 0.001 $\mu$m and a maximum of $a_{\rm max}$ = 10 $\cdot$ $a_{\rm eff}$ $\mu$m, where $a_{\rm eff}$ is the effective cloud particle radius (Eq. \ref{eq:aeff}).
During the simulation, by sampling a random number $\zeta$ $\in$ (0,1), the particle size from the distribution interacting with the packet can therefore be stochastically determined by sampling  $A_{\rm CDF}$ (i.e. $\zeta$ $\rightarrow$ $a$).

We assume two scattering regime limits in our model for a certain cloud particle size parameter ($x = 2\pi a /\lambda$).
For x $<$ 0.1 we assume a Rayleigh scattering event, while if x $\ge$ 0.1 we assume a TTHG scattering event.
We assume TTHG scattering events to occur at the properties of the effective mean radius values (e.g.  $g_{\lambda}$($a_{\rm eff}$), Sect. \ref{sec:gdopac}).
The effect of this scheme is that if the mean and variance skew the area distribution towards larger cloud particle sizes, then the fractional contribution of particle areas, where x $\ge$ 0.1 increases, and so the probability of a TTHG event is increased.
However, if the distribution is skewed towards smaller cloud particle sizes then a greater fraction of cloud particle area satisfies x $<$ 0.1, increasing the likelihood of a Rayleigh event.

\subsection{MCRT variance reduction techniques}
\label{sec:var_red}

\begin{figure*}
   \centering
   \includegraphics[width=0.49\textwidth]{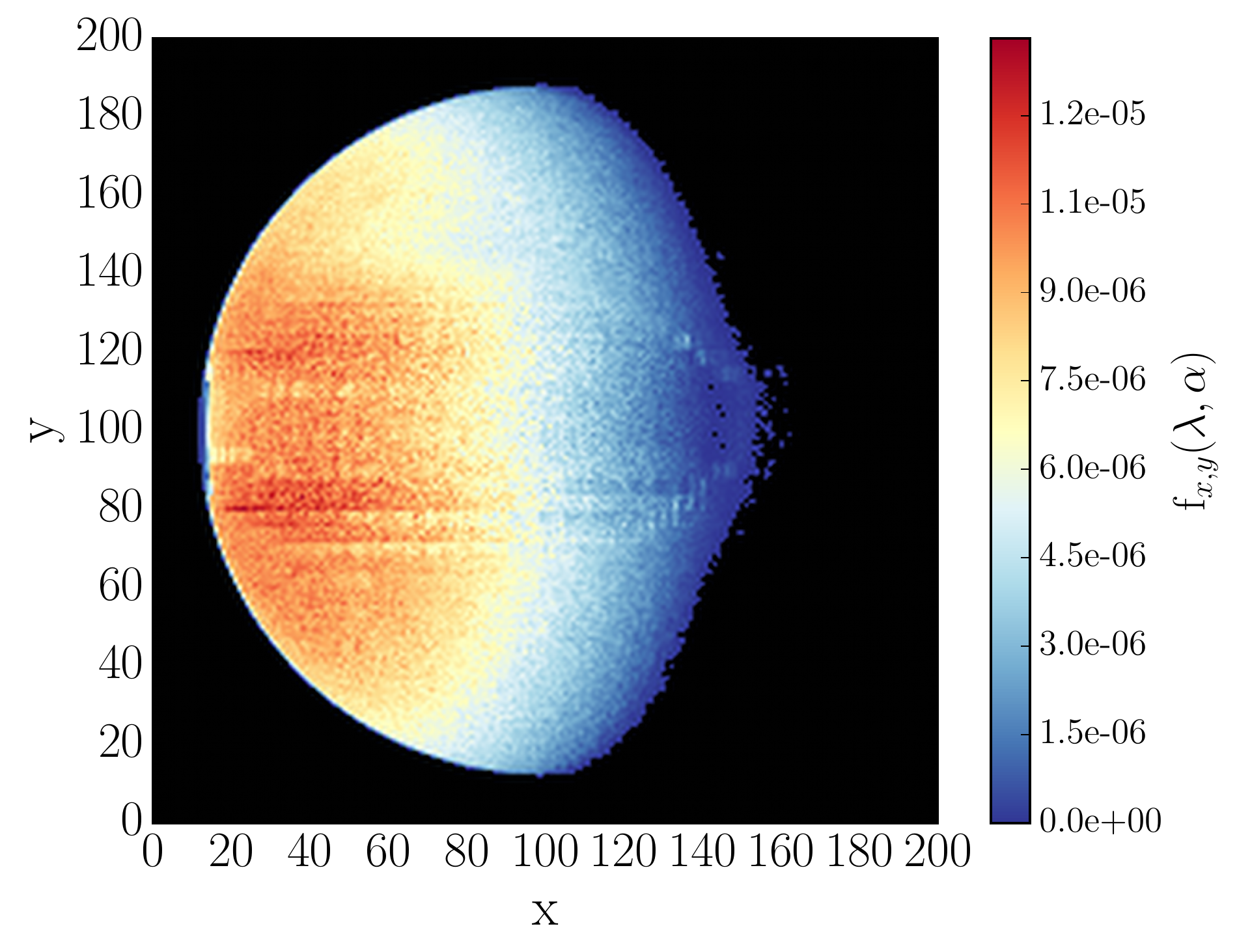} 
   \includegraphics[width=0.49\textwidth]{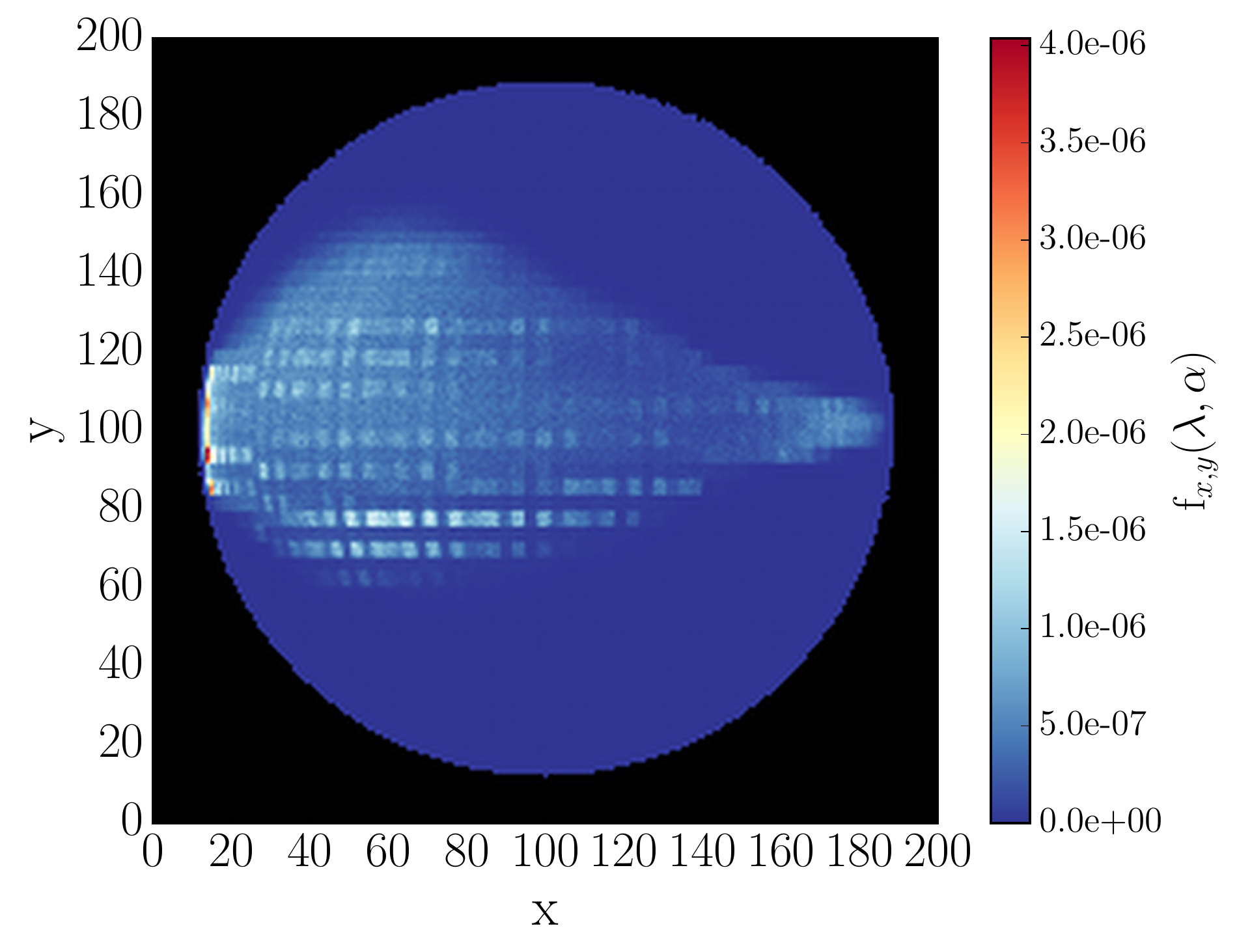}
   \caption{Incident scattered (left) and atmospheric emitted (right) light image array of the L-packet luminosity fractions f$_{x,y}$($\lambda, \alpha$) $\in$ (0,1) escaping from the HD 189733b simulation.
   These images are produced by the next event estimator method at 1.0 $\mu$m for a viewing angle/phase of $\alpha$ = 60\degr (eastern dayside).
   The sharp transition from blue to black at pixel numbers x = 120$-$150 is the eastern terminator in the scattered light image, including a "twilight" effect where packets scatter past the terminator line.
   The effect of the grid structure on the images is more apparent in the emitted light, since packets are initialised from within the cell volumes.}
   \label{fig:peeloff_example}
\end{figure*}

In a basic MCRT scheme, to produce observable quantities for a particular observation direction, the MCRT code would track how many packets of a particular luminosity entered the simulation and how many escaped.
The ratio of these values would then give the fraction of luminosity escaping the simulation domain towards a particular direction.
However, the probability of an individual packet escaping by chance towards a particular direction can be incredibly small, depending on the geometric properties of the simulation, and can produce noisy results if not enough Monte Carlo packets are simulated \citep[e.g. see][]{Dupree2002}.
Therefore, variance reduction techniques were developed for MCRT, which allows for a much improved signal to noise and greater computational efficiency than the basic scheme.
We employ three techniques in this study as follows: next event estimation, survival biasing, and composite emission biasing.
The variance and convergence properties our model output is presented in Sect. \ref{sec:var}.

\subsubsection{Next event estimation}
\label{sec:peelingoff}
Next event estimation \citep[e.g.][]{Yusef-Zadeh1984, Wood1999} is a variance reduction method that uses a ray-tracing technique in conjunction with the Monte Carlo scheme.
At every scattering or emission event, a fraction of the packet's luminosity, W$_{\rm po}$$\epsilon_{0}$, (Eq. \ref{eq:photon_energy}) is peeled off towards an observation direction using a ray-tracing method.
The fraction of luminosity is given by the normalised phase function of the type of interaction, weighted by the probability of the packet escaping the simulation towards this observation direction.
The peeled off luminosity is wavelength dependent and the luminosity carried by the packet is not altered in the peeling off scheme.
The peeled off luminosity is then stored in a 2D pixel image array at this observation direction, which can be used to derive observational quantities.

Multiple pixel images can be constructed for many viewing angles and wavelengths, which allows a single photon packet interaction to contribute to multiple output images at different planetary phases.
Since every packet now contributes to the output images, this improves the signal to noise of the output as images can be produced without relying on packets emerging by chance from the simulation boundaries towards the observer.
Figure \ref{fig:peeloff_example} illustrates the resultant next event estimator scattered stellar (left) and atmospheric emitted (right) luminosity fractions  at 1.0 $\mu$m for a viewing angle of $\phi$ = 60\degr, $\theta$ = 0\degr.

For atmospheric emission, assumed to be isotropic, the fraction of the packet luminosity peeled off towards observation direction $\alpha$ is given by
\begin{equation}
W_{\rm po}(\alpha) = \frac{1}{4\pi}\rm{e}^{-\tau_{2}},
\label{eq:w_em}
\end{equation}
where 1/4$\pi$ is the normalisation factor for isotropic emission into 4$\pi$ steradians and $\tau_{2}$($\lambda$) (where indexed $\tau$ terms (e.g. $\tau_{1}$,$\tau_{2}$ etc.) indicate optical depths calculated using the ray-tracing method to the simulation boundaries, and non-indexed $\tau$ denotes the stochastically sampled optical depth for photon packets) the optical depth through the computational zone at wavelength $\lambda$, towards the observational direction.

For non-isotropic scattering events such as Rayleigh or Henyey-Greenstein scattering, the fraction of peeled off luminosity is related to the probability of the packet scattering towards the observational direction.
For Rayleigh scattering, the weight of peeled off luminosity is given by
\begin{equation}
W_{\rm po}(\alpha) = \Phi_{\rm RS}(\alpha) \rm{e}^{-\tau_{2}},
\end{equation}
and for Two-term Henyey-Greenstein, 
\begin{equation}
W_{\rm po}(\alpha) = \Phi_{\rm TTHG}(g_{a},g_{b}, \alpha)\rm{e}^{-\tau_{2}}.
\end{equation}
Phase curve information can be calculated by adding the contribution of each event to an image in the desired observation direction (e.g.. 360\degr\ in longitude).
Each weight is added into a square image pixel array of (x, y) = (201, 201) pixels and then normalised by the number of initialised packets after the simulation is complete.
The luminosity scattered/emitted in the observational direction can then be found by multiplying the fraction of escaping luminosity by the total luminosity of the emitting source.

\subsubsection{Survival biasing}
\label{sec:rrw}

For diffuse emission from within the simulation boundaries, the packet is forced to scatter at each interaction and not terminated upon absorption (see Sect. \ref{sec:Method}).
In order to conserve energy, the weight of the luminosity carried by the packet is reduced at each interaction by the single scattering albedo
\begin{equation}
W_{\rm new} = \omega W,
\end{equation}
where, in an interaction with the gas phase, $\omega$ = $\omega_{\rm gas}$ (Eq. \ref{eq:gas_alb}) and with the cloud particles  $\omega$ = $\omega_{\rm cloud}$ (Eq. \ref{eq:cloud_alb}).
The fraction of luminosity lost by the photon packet due to absorption is then taken into account in the peeled off images as the packet interacts with the surrounding medium.
So that a packet does not scatter indefinitely with ever decreasing weight, a Russian Roulette scheme \citep[e.g.][]{Dupree2002} is applied to stochastically terminate packets below a predefined weight cut-off.
The packet is then given a 1 in 10 chance of surviving, with the new weight of the surviving packets W$_{\rm new}$  = 10 $\cdot$ W.
In our model, packets with weight W $<$ 10$^{-3}$ are entered into the Russian Roulette scheme.

\subsubsection{Composite emission biasing}
\label{sec:comp_bias}

For hot Jupiter atmospheres, the total dayside emission luminosity can be orders of magnitude greater than the emission from the nightside of the planet.
In non-biased MCRT, the number of packets $N_{i, \lambda}$ emitted from cell $i$ is given by the fraction of the luminosity of the cell to the total luminosity of the atmospheric cells at this wavelength \citep[e.g.][]{Pinte2006},
\begin{equation}
\label{eq:num_source}
N_{i, \lambda} = \frac{L_{i, \lambda}}{\sum_{i} L_{i, \lambda}} N_{\rm atm},
\end{equation}
where $L_{i, \lambda}$ is the luminosity of the cell, $\sum_{i} L_{i, \lambda}$ the total luminosity of the considered cells, and $N_{\rm atm}$ the total number of atmospheric L-packets emitted at wavelength $\lambda$.
Low luminosity regions (e.g.. the nightside), where $L_{i, \lambda} \ll \sum_{i} L_{i, \lambda}$  can therefore be under sampled in the emission scheme, which increases the variance of results derived from these regions.

To address this, we implement the multiple component, emission composite biasing method from \citet{Baes2016}, where a linear combination of the unbiased emission (Eq. \ref{eq:num_source}) probability distribution function and a uniform (in cell number) distribution function is applied.
This results in the number of emitted packets per cell given by
\begin{equation}
\label{eq:num_bias}
N_{i, \lambda} = N_{\rm atm} \left[(1 - \eta)\frac{L_{i, \lambda}}{\sum_{i} L_{i, \lambda}} + \frac{\eta}{\sum_{i} i_{\lambda}}\right],
\end{equation}
where $\eta$ $\in$ (0,1), is a parameter governing the linear combination and $\sum_{i} i_{\lambda}$ the total number emitting cells at wavelength $\lambda$.
To conserve the total luminosity, each emitted packet is given a cell dependent starting weight of 
\begin{equation}
W_{\rm em} = \frac{1}{(1 - \eta) + \eta\langle L_{i, \lambda} \rangle / L_{i, \lambda} },
\end{equation}
where 
\begin{equation}
\langle L_{i, \lambda} \rangle  = \frac{{\sum_{i} L_{i, \lambda}}}{\sum_{i} i_{\lambda}},
\end{equation}
is the average cell luminosity at wavelength $\lambda$.
We apply a fixed $\eta$ = 0.5 throughout our simulations.
\citet{Baes2016} provide an in depth description of this composite emission biasing method.

\section{Input and output quantities}
\label{sec:Method}

In this section, we provide details of our input quantities to the MCRT scheme and the relationships between the MCRT output, which is in normalised luminosity fraction units, and observational quantities (geometric albedos, flux ratios, etc.).
We perform all scattering and emission calculations for 250 wavelengths in the range 0.3 and 5.0 $\mu$m, linear in wave-number space.
Table \ref{tab:sys_prop} shows the HD 189733 star and planet parameters used in the simulation.

\begin{table}[htdp]
\caption{HD 189733 system parameters adopted for this study. We use the parameters presented in \citet{Torres2008}.}
\begin{center}
\begin{tabular}{|c|c|}
Parameter & Value \\ \hline
R$_{\star}$ & 0.76 R$_{\odot}$ \\
T$_{\rm eff, \star}$ & 5040 K \\
a & 0.031 au \\
R$_{p}$ & 1.138 R$_{\rm Jup}$ \\
\end{tabular}
\end{center}
\label{tab:sys_prop}
\end{table}%

\subsection{Summary of 3D RHD results}
We briefly provide a summary of the results of our 3D cloud forming RHD simulation of HD 189733b (Paper I), which are used as input to the MCRT post-processing simulation.
The mean cloud particle sizes in the upper atmosphere (0.1 - 100 mbar) range from seed particle sizes $\langle$a$\rangle$ $\sim$ 0.001 $\mu$m at the hottest regions of the dayside to  $\langle$a$\rangle$ $\sim$ 0.1 $\mu$m at higher latitudes and the cooler nightside.
Cloud particles to the east of the sub-stellar point are generally smaller than those to the west of the sub-stellar point, with $\sim$1 order of magnitude differences between cloud particle sizes at the east and west terminator regions.
Cloud particle composition also varies with latitude and longitude; TiO$_{2}$[s] rich grains dominate the hotter dayside equatorial regions, while silicate materials (SiO[s], SiO$_{2}$[s], MgSiO$_{3}$[s], Mg$_{2}$SiO$_{4}$[s]) dominate the cooler higher latitudes and nightside regions.
TiO$_{2}$[s] rich grains are also more prevalent on the eastern terminator region, while silicates dominate the western regions. 
This material composition disparity, due to differences in the thermal stability between TiO$_{2}$[s] and silicates, is the main reason why each terminator region contains different cloud particle sizes.

For the MCRT post-processing simulation, we use the same spherical RHD simulation dimensions of radial, longitude and latitude sizes  (R, $\phi$, $\theta$)  = (100, 160, 64) respectively.
Latitude $\theta$ cells defined near the pole edges in the RHD simulation are assumed to extend to $-\pi$ and $\pi$ radians respectfully.
This is required so that no regions remain undefined across the whole sphere in the MCRT simulation.
Cell thermodynamic quantities (T$_{\rm gas}$, $\rho_{\rm gas}$) and cloud moment properties (L$_{\rm j}$) are taken from the snapshot ($\sim$ 60 Earth days) simulation of Paper I.
The cell thermodynamic states are kept constant at the values given by the RHD model throughout the MCRT simulations.

\subsection{Cloud and gas opacity}
\label{sec:gdopac}

The input opacities are critical to the accuracy of the MCRT scheme as they govern the distance travelled by the stochastically sampled $\tau_{\lambda}$, the probability of packets scattering, and the type of scattering events.
The total extinction opacity $\kappa_{\rm ext,total}$ [cm$^{2}$ g$^{-1}$] in each cell is given by
\begin{equation}
\label{eq:ktot}
 \kappa_{\rm ext, total} = \kappa_{\rm abs, gas}  + \kappa_{\rm sca, H_{2}}  + \kappa_{\rm ext, cloud}, 
\end{equation}
where $\kappa_{\rm abs, gas}$ [cm$^{2}$ g$^{-1}$] is the absorption due to gaseous atoms and molecules, $\kappa_{\rm sca, H_{2}}$ [cm$^{2}$ g$^{-1}$] is the contribution of Rayleigh scattering by molecular H$_{2}$, and $\kappa_{\rm ext, cloud}$ = $\kappa_{\rm abs, cloud}$ + $\kappa_{\rm scat, cloud}$ is the total extinction (absorption + scattering) due to cloud particles.

For gas absorption opacities $\kappa_{\rm abs, gas}$, we adopt the solar metallicity, chemical equilibrium \citet{Sharp2007} opacity tables, without TiO and ViO opacities, linearly interpolated to the RHD cell thermodynamic properties.
We perform k-distribution \citep[e.g.][]{Goody1989,Lacis1991,Grimm2015} band averages from the high-resolution opacity tables across 250 bins, linear in wave-number between (wavelengths) 0.3 and 5 $\mu$m.
Eight Gaussian quadrature points were used with a split at 95\% of the cumulative opacity, in the same manner as \citet{Showman2009}.

For the H$_{2}$ scattering opacity we use the cross-section, wavelength relation from \citet{Dalgarno1962} given by
\begin{equation}
\sigma_{sca, H_{2}}(\lambda) = \frac{8.14 \times 10^{-13}}{\lambda^{4}} + \frac{1.28 \times 10^{-6}}{\lambda^{6}} + \frac{1.61}{\lambda^{8}} ,
\end{equation}
where $\lambda$ is in \angstrom\ and $\sigma_{sca, H_{2}}$ in cm$^{2}$.
This is converted into a mass opacity by the relation $\kappa_{sca, H_{2}}$($\lambda$) = $n_{H_{2}}$$\sigma_{sca, H_{2}}(\lambda)$ / $\rho_{\rm gas}$.
The abundance of molecular H$_{2}$ ($n_{H_{2}}$ [cm$^{-3}$])  is calculated at each cell assuming chemical equilibrium using the routines from \citet{Helling2016}.

For cloud particle absorption and scattering opacities we calculate a mono-disperse absorption $\kappa_{\rm abs, cloud}$ and scattering $\kappa_{\rm sca, cloud}$ opacity of the cloud particles in each cell given by
\begin{eqnarray}
\label{eq:cloud_opc}
\kappa_{\rm abs, cloud}(\lambda) = \pi a^{2} Q_{\rm abs}(\lambda, a) n_{\rm d} / \rho_{\rm gas} , \nonumber \\
\kappa_{\rm sca, cloud}(\lambda) = \pi a^{2} Q_{\rm sca}(\lambda, a) n_{\rm d} / \rho_{\rm gas} ,
\end{eqnarray}
respectively, where $a$ [cm] is cloud particle size, $Q_{\rm abs}$ and $Q_{\rm sca}$ are the absorption and scattering efficiency factors given by the \textsc{MieX} code of \citet{Wolf2004}, and $n_{\rm d}$ [cm$^{-3}$] is the total cloud particle number density.
We calculate the cloud opacities at the effective grain size, $a_{\rm  eff}$, which is given by the ratio of the third (L$_{3}$) and second (L$_{2}$) cloud moments \citep{Hansen1974}
\begin{equation}
\label{eq:aeff}
a_{\rm eff} = \sqrt[3]{\frac{3}{4\pi}} \frac{L_{3}}{L_{2}}  ,
\end{equation}
which is the area weighted average cloud particle size across the grain size distribution.
The single scattering albedo $\omega_{\rm cloud}$ (Eq. \ref{eq:cloud_alb}) and scattering asymmetry parameter $g_{\lambda}$ (Eq. \ref{eq:mean_cosine}) are also calculated at the effective cloud particle size.
Effective (n, k) coefficients of the cloud particles required as input for Mie theory are calculated for mixed material cloud particles using effective medium theory in the same way as \citet{Lee2015b}.

\subsection{Scattered light phase curves and albedo spectra}

\begin{figure}
   \centering
   \includegraphics[width=0.49\textwidth]{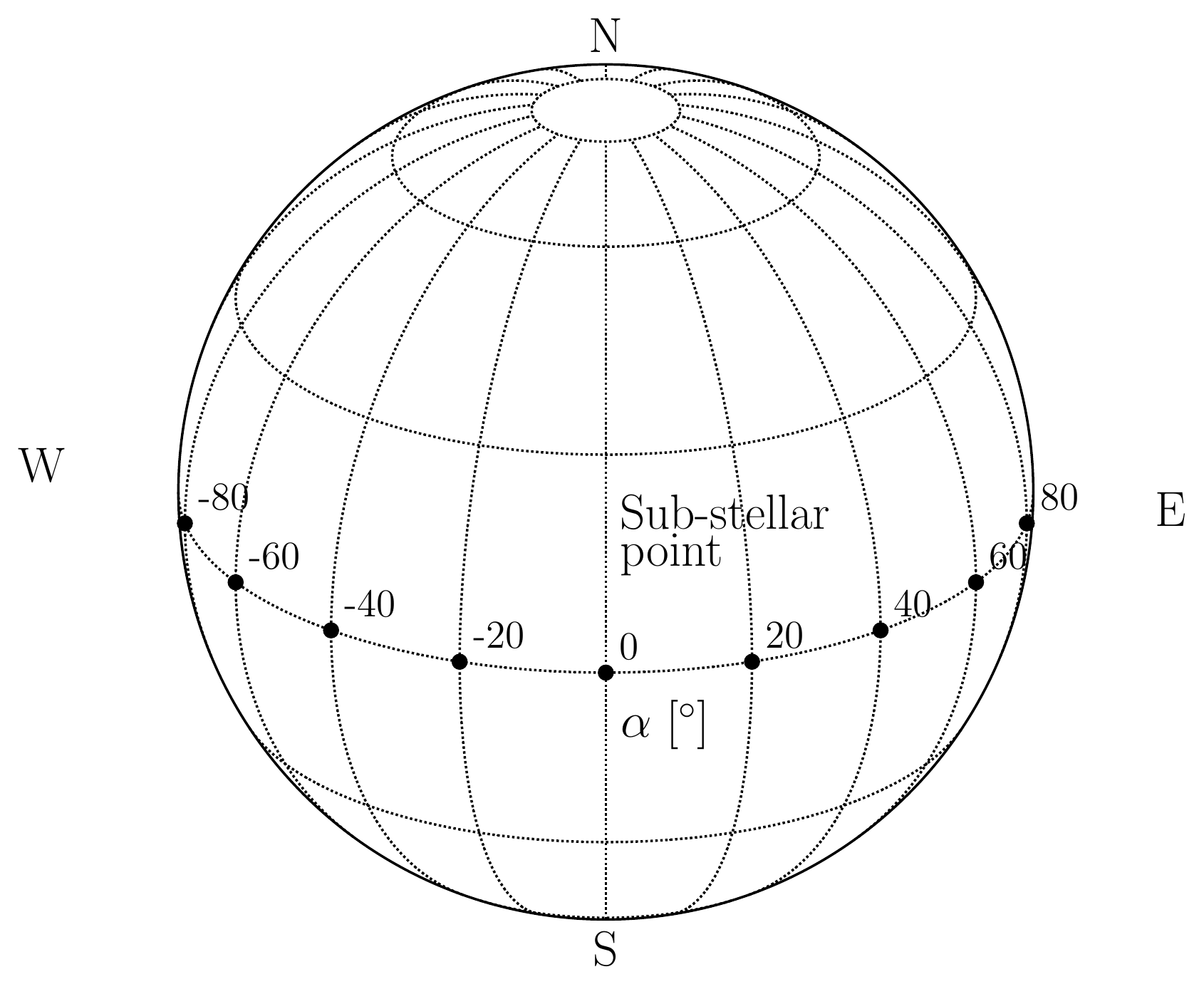} 
   \caption{Diagram visualising the definition of equatorial viewing angle $\alpha$ [\degr] used in this study. Viewing angles east of the sub-stellar point are positively defined, while regions west of the sub-stellar point are negatively defined.}
   \label{fig:alpha_def}
\end{figure}

To produce scattered light phase curves, we track the random walks of stellar L-packets incident on the HD 189733b simulated atmosphere.
The stellar illumination is assumed to be plane parallel from the direction of the host star, with the initial latitude and longitude ($\phi$, $\theta$) top of atmosphere positions determined stochastically from a rejection method, carried out across the circular annulus of the dayside face of the planet.
In this mode, should the packet be absorbed by the gas or cloud particles, it is terminated and no longer contributes to the simulation.
The survival biasing variance reduction technique is not required in this mode since the majority of incident stellar packets are scattered or absorbed in the atmosphere.
Therefore, the fractional weight of the packet luminosity remains unchanged in this mode (W = 1).
We chose a viewing angle of $\theta$ = 0\degr\ latitude (equator) while altering the longitude viewing angle $\alpha$ to capture planetary phase information.
We define $\alpha$ = 0\degr\ as the viewing angle at the sub-stellar point (i.e.. full phase) with $+$\degr $\alpha$ eastward and $-$\degr  $\alpha$ westward  from the sub-stellar point.
Figure \ref{fig:alpha_def} shows a schematic visualisation of this $\alpha$ definition.
As a consequence, we ignore possible viewing effects of non-negligible transit impact parameters, similar to other post-processing RHD/GCM methods \citep[e.g.][]{Fortney2006,Showman2009,Amundsen2016}.

The total fraction of luminosity, carried by the photon packets, escaping towards viewing angle $\alpha$ at wavelength $\lambda$ is given by summing the image pixel array produced by the peeling off method (Sect. \ref{sec:peelingoff}, Fig. \ref{fig:peeloff_example})
\begin{equation}
f_{\rm total}(\lambda, \alpha) = \sum_{x,y} f_{x,y}(\lambda, \alpha),
\end{equation}
where $f_{\rm total}$ is the total fraction of luminosity escaping towards viewing angle $\alpha$ at wavelength $\lambda$ and $f_{x,y}$ is the fraction of luminosity contained in a (x,y) pixel image array.
The monochromatic phase dependent albedo A$_{\lambda}$($\alpha$) is defined as
\begin{equation}
\label{eq:albspec}
A_{\lambda}(\alpha) = \frac{f_{total}(\lambda, \alpha) L_{inc}(\lambda)}{L_{inc}(\lambda)} = f_{total}(\lambda, \alpha), 
\end{equation}
where L$_{inc}$ [erg s$^{-1}$] is the incident luminosity onto the dayside face of the exoplanet atmosphere from the host star. 
The monochromatic apparent geometric albedo A$_{g, \lambda}$ is defined as the fraction of scattered light at zero phase angle ($\alpha$ = 0\degr) to an equivalent spherical Lambertain surface \citep[e.g..][]{Seager2010,Madhusudhan2012}.
In the MCRT scheme A$_{g,\lambda}$ is derived from the luminosity fractions by the equation
\begin{equation}
\label{eq:geoalb}
A_{g,\lambda} = \frac{f_{total}(\lambda, \alpha = 0\degr)L_{inc}(\lambda)}{f_{Lambert}L_{inc}(\lambda)} = \frac{f_{total}(\lambda, \alpha = 0\degr)}{2/3},
\end{equation}
where $f_{Lambert}$ = 2/3 is the scattering fraction of the theoretical Lambertian planet at zero phase.

The wavelength-dependent classical phase function \citep[e.g.][]{Seager2010,Madhusudhan2012} can be constructed from the scattering fractions
\begin{equation}
\label{eq:phasefunction}
\Phi_{\lambda}(\alpha) = \frac{f_{\rm total}(\lambda, \alpha)}{f_{\rm total}(\lambda, \alpha = 0\degr)},
\end{equation}
which is the ratio of reflected luminosity at viewing angle $\alpha$ compared to the fraction reflected at $\alpha$ = 0\degr.
The monochromatic planetary luminosity due to reflected starlight L$_{\rm p, scat, \lambda}$ [erg s$^{-1}$ cm$^{-1}$] as a function of viewing angle is given by
\begin{equation}
L_{\rm p, scat, \lambda}(\alpha)  = f_{\rm total}(\lambda, \alpha) L_{\star, \lambda} \left(\frac{R_{p}^{2}}{4a^{2}}\right)  , 
\end{equation}
where $R_{p}$ is the radius of the planet and $a$ the semi-major axis.
The monochromatic luminosity of the star is given by
\begin{equation}
\label{eq:star_lum}
L_{\star, \lambda} = 4 \pi^{2}R_{\star}^{2}B_{\lambda} (T_{\rm \star, eff}, \lambda), 
\end{equation}
where $R_{\star}$ is the radius of the star and $B_{\lambda}$ the Planck function, which is dependent on the stellar effective temperature $T_{\rm \star, eff}$.
To ensure good signal to noise, we emit the same number of packets (N$_{\rm inc}$ = 10$^{7}$) for each (pseudo) monochromatic wavelength.

\subsection{Emitted light phase curves and spectra}

\begin{figure*}
   \centering
   \includegraphics[width=0.49\textwidth]{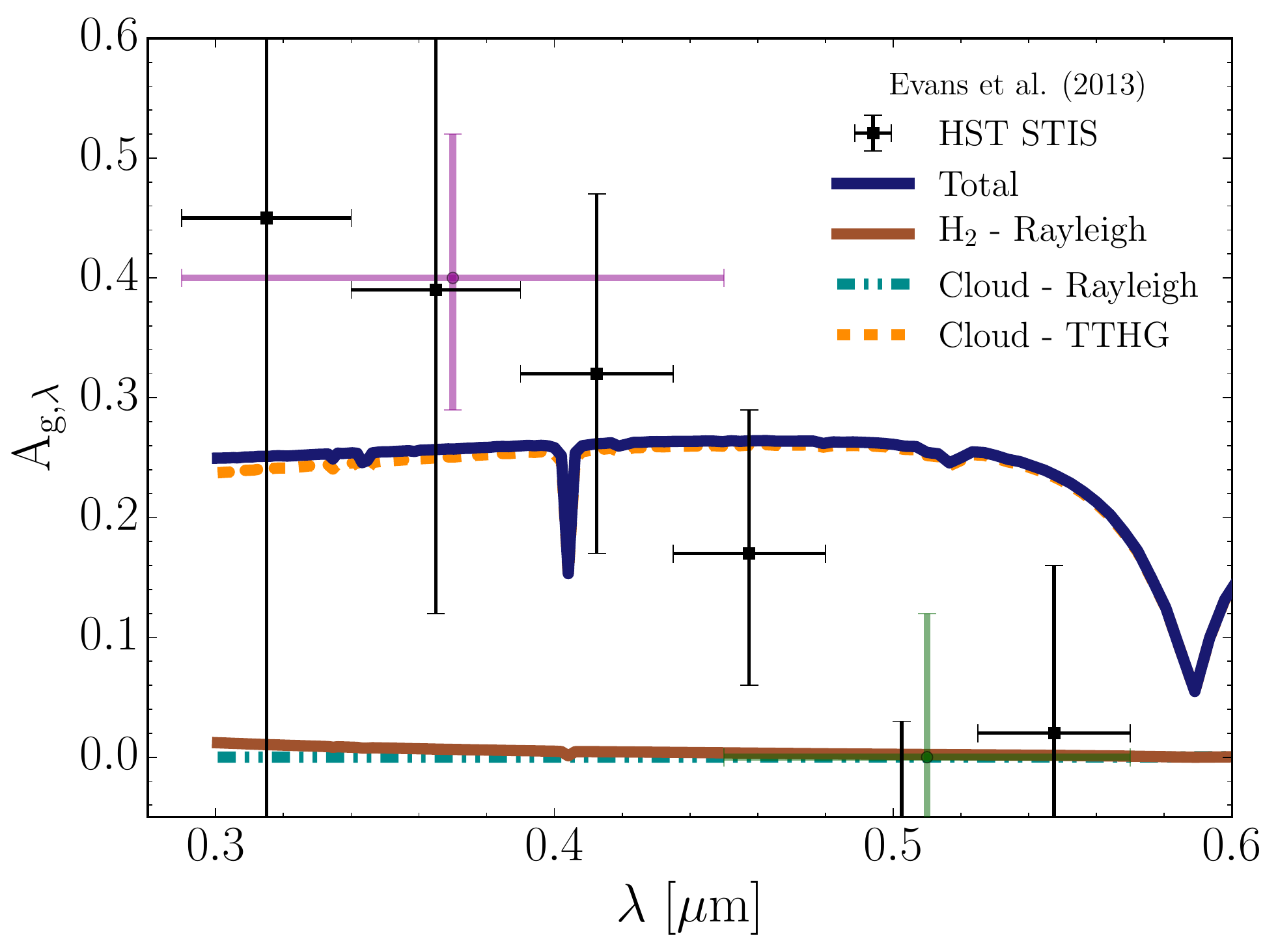}   
   \includegraphics[width=0.49\textwidth]{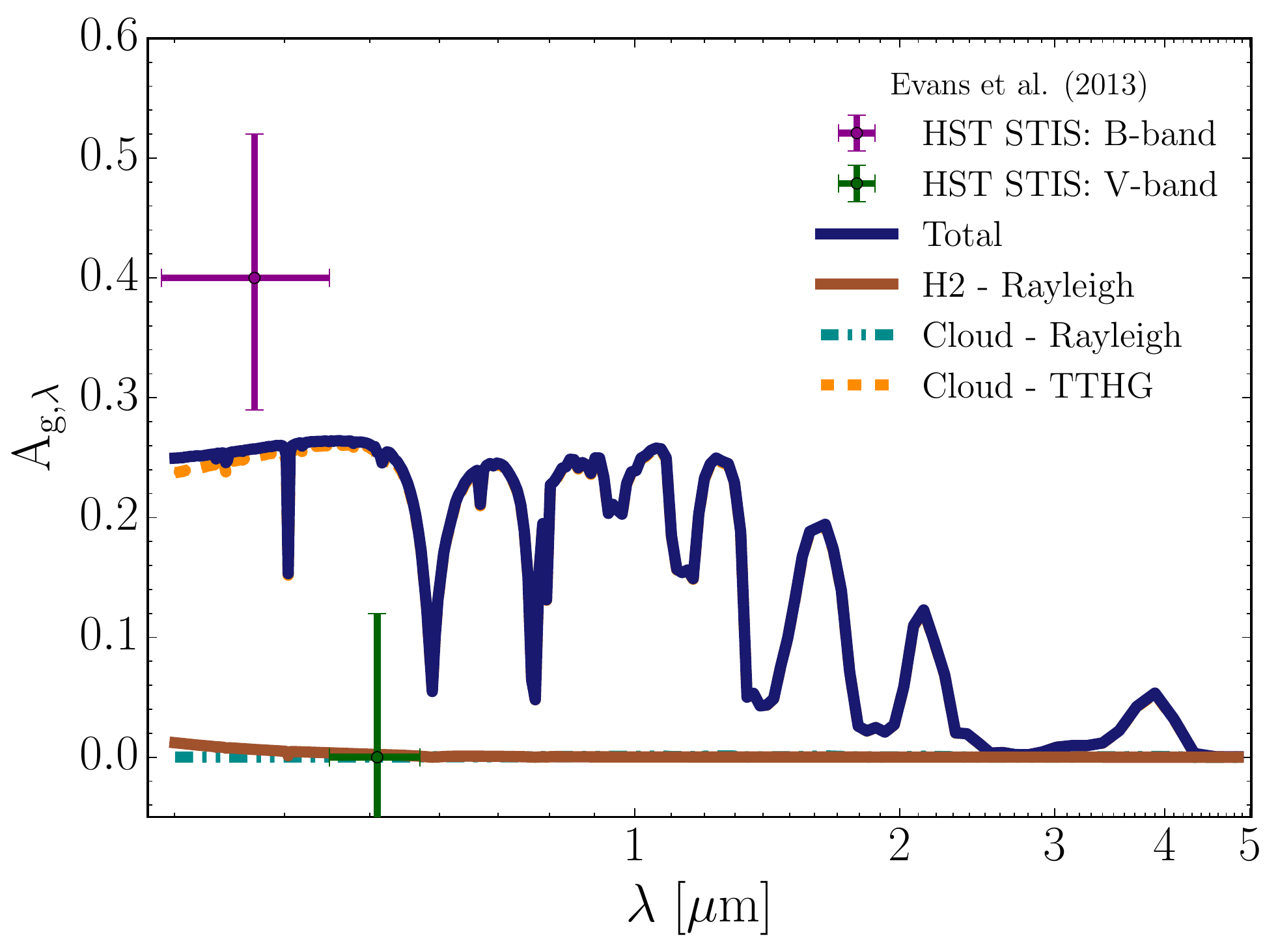}
   \caption{Scattered light apparent geometric albedo A$_{g, \lambda}$ (Eq. \ref{eq:geoalb}) of our HD 189733b simulation output compared to \citet{Evans2013}'s individual HST STIS measurements.
   The total contribution is indicated by the solid blue line; TTHG cloud particle scattering by the dashed orange line; Rayleigh cloud particle scattering by the teal dash-dotted line; H$_{2}$ Rayleigh scattering by the brown solid line.
   Left: (0.29-0.6 $\mu$m) The Rayleigh \& TTHG model is generally consistent with the B Band HST STIS measurements, but does not predict the low albedo of the V band.
   Right: (0.29-5 $\mu$m) Extended spectra from 0.3-5 $\mu$m, showing the convolved B Band and V band measurements from \citet{Evans2013}. 
   A$_{g, \lambda}$ becomes negligible beyond 5 $\mu$m.}
   \label{fig:geoalb}
\end{figure*}

We follow a similar scheme to the protoplanetary disk and ISM MCRT based studies of \citet{Pinte2006} and \citet{Robitaille2011} where the monochromatic luminosity L$_{i, \lambda}$ [erg s$^{-1}$ cm$^{-1}$]  of computational cell $i$ is 
\begin{equation}
L_{i, \lambda} = 4\pi\rho_{i, gas}V_{i}(\kappa_{i, \lambda, abs, cloud} + \kappa_{i, \lambda, abs, gas})B_{\lambda} (T_{i}) , 
\end{equation}
where $V_{i}$ is the volume of the cell, $\rho_{i, gas}$ the gas density, $\kappa_{i, \lambda, abs, cloud}$ the cloud absorption opacity, $\kappa_{i, \lambda, abs, gas}$ the gas absorption opacity, and $B_{i, \lambda}$ the Planck function, which is dependent on the cell temperature $T_{i}$.

In this scheme the packets are not terminated upon absorption but are forced to scatter with reduction in their weight at each event.
Survival biasing and Russian Roulette techniques are therefore applied (Sect. \ref{sec:rrw}).
Composite emission biasing is also applied to decrease the noise error of results derived from nightside regions (Sect. \ref{sec:comp_bias}).
We employ a wavelength-dependent \textit{dark zone} \citep{Pinte2006} of longitude- and latitude-dependent, radially integrated optical depth $\tau_{r, \lambda}(\theta, \phi)$ = 30 (Eq. \ref{eq:tau}).
No packets are emitted from these regions and packets are terminated if they cross into regions deeper in the atmosphere than the cell radial index given by this condition.
This prevents tracking packets through low luminosity weight regions which contribute negligibly to the photospheric emission.
The contribution of cells below this depth to the total cell emission luminosity is also ignored (i.e. $L_{i, \lambda}$ = 0).
The emitted monochromatic luminosity L$_{\rm p, em, \lambda}$ [erg s$^{-1}$ cm$^{-1}$] from the planet at viewing angle $\alpha$ is then given by 

\begin{equation}
\label{eq:emitted}
L_{\rm p, em, \lambda}(\alpha) =  f_{\rm total}(\lambda, \alpha) \sum_{i} L_{i, \lambda}.
\end{equation}
As in the scattered light case, we assume the same number of packets (N$_{\rm atm}$ = 10$^{8}$) for  each (pseudo) monochromatic wavelength.

\section{Results}
\label{sec:Results}

In this section, we present the scattered and emitted light results of our MCRT post-processing method.
We chose 36 different viewing angles to capture orbital phase information of the scattered and emitted light, which gives a phase resolution of $\Delta\alpha$ = 10\degr.
Sections \ref{sec:geoalb} and \ref {sec:scatcurves} present our incident scattered light apparent geometric albedo A$_{g}$, and albedo spectra/phase curves, respectively. 
Section \ref{sec:emitted} presents our atmospheric emitted luminosity spectra and phase curves.
In Sect. \ref{sec:res_combine} we combine our scattered and emitted light results and compare these combined results with current observations; we also predict Kepler, TESS and CHEOPS photometric observations for HD 189733b.
Finally, in Sect. \ref{sec:var} we discuss the variance and convergence properties of our MCRT model results.

\subsection{Geometric albedo}
\label{sec:geoalb}

Figure \ref{fig:geoalb} shows the resultant geometric albedo of our post-processed cloud forming HD 189733b 3D RHD model, which we compare to the HST STIS observations of \citet{Evans2013}.
Our results in the B band are consistent with the individual STIS data points, but underestimate the convolved B Band geometric albedo.
Our results are also consistent with the \citet{Wiktotowicz2015} ground based B+V band polarimetry measurement of an upper limit of A$_{g, \lambda}$ $<$ 0.4.
Our V Band result compares poorer to the observations, however it is clear that Na absorption is responsible for the downward trend near 0.6 $\mu$m.
This is perhaps because of the presence of an unknown absorbing material in the V band (Sect. \ref{sec:Discussion}).

Figure \ref{fig:geoalb} also shows the individual contributions from H$_{2}$ Rayleigh scattering, cloud particle Rayleigh scattering, and cloud particle TTHG scattering to the total fractions.
The TTHG scattering contributes the greatest fraction to the total geometric albedo across all wavelengths.
A small H$_{2}$ Rayleigh scattering component is also present at shorter optical wavelengths.

\subsection{Scattered light phase curves}
\label{sec:scatcurves}

During a single orbit of HD 189733b around its parent star, different atmospheric regions come in and out of view of the detector, dependent on the planetary phase being observed.
If eastward and westward hemispheres of the dayside atmospheric properties are different, an asymmetric signal as a function of phase ($\alpha$) is expected to be observed.
In order to extract this phase behaviour from our 3D RHD results, we produce observables at various viewing angles (longitude from the sub-stellar point) $\alpha$ during the Monte Carlo simulation.

\begin{figure*}
   \centering
   \includegraphics[width=0.75\textwidth]{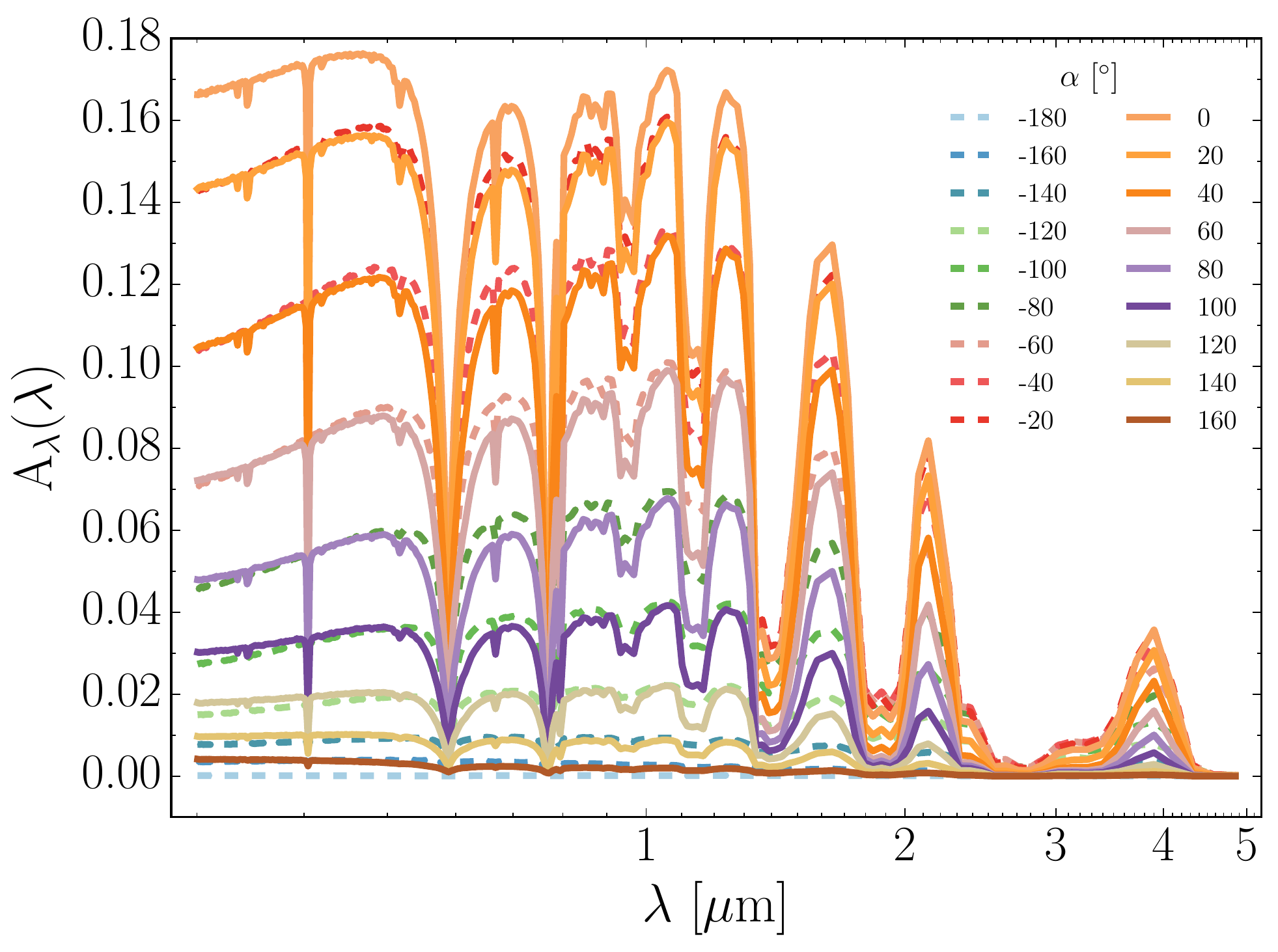} 
   \caption{Scattered light albedo spectra A$_{\lambda}$($\alpha$) (Eq. \ref{eq:albspec})  as a function of wavelength at several viewing angles $\alpha$, where $\alpha$ = 0\degr defines the sub-stellar point.
   Differences in the 3D cloud structure on the east and west hemispheres from the sub-stellar points produce small variations in A$_{\lambda}$ across the 360\degr\ phase space at optical wavelengths.
   Na and K absorption are responsible for the drops in A$_{\lambda}$ in the optical, while H$_{2}$O features are the main absorber at near-IR to IR wavelengths.}
   \label{fig:Ophase_spec}
\end{figure*}

Figure \ref{fig:Ophase_spec} shows the albedo A$_{\lambda}$($\alpha$) spectra (Eq. \ref{eq:albspec}) as a function of phase.
Slight differences in the fraction of scattered light between the regions east and west of the sub-stellar point are present, due to the differences in the east-west 3D cloud properties in our HD 189733b RHD model.
The greatest difference in A$_{\lambda}$ ($\Delta$ A$_{\lambda}$ = 0.005) in the B and V bands between the east and west hemispheres occur between viewing angles of $\sim$ 80\degr and 100\degr.
This corresponds to the east and west day-night terminator regions, respectively, which have the largest differences in cloud properties.
For example, the eastern terminator generally has mean particles sizes $\langle$a$\rangle$ $\sim$ 0.01 $\mu$m, composed of a mix of Si-O materials and TiO$_{2}$.
While the western terminator $\langle$a$\rangle$ $\sim$ 0.1 $\mu$m with larger volume fractions of MgSiO$_{3}$[s] and Mg$_{2}$SiO$_{4}$[s] (Paper I).
Wavelengths $>$ 5 $\mu$m show little or no scattering behaviour, as the single scattering albedo for both gas and cloud components become negligible.

\begin{figure*}
   \centering
   \includegraphics[width=0.49\textwidth]{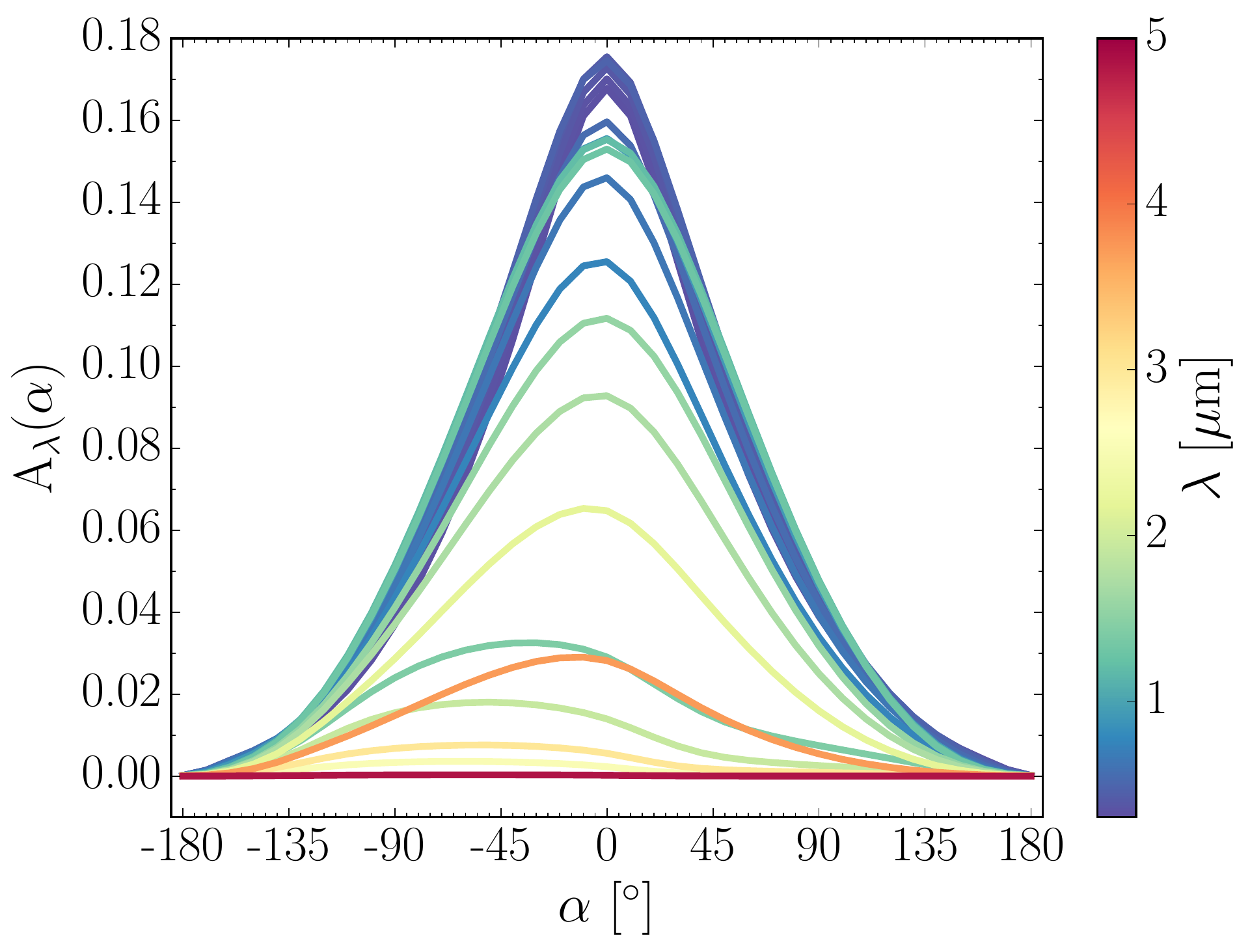}       
   \includegraphics[width=0.49\textwidth]{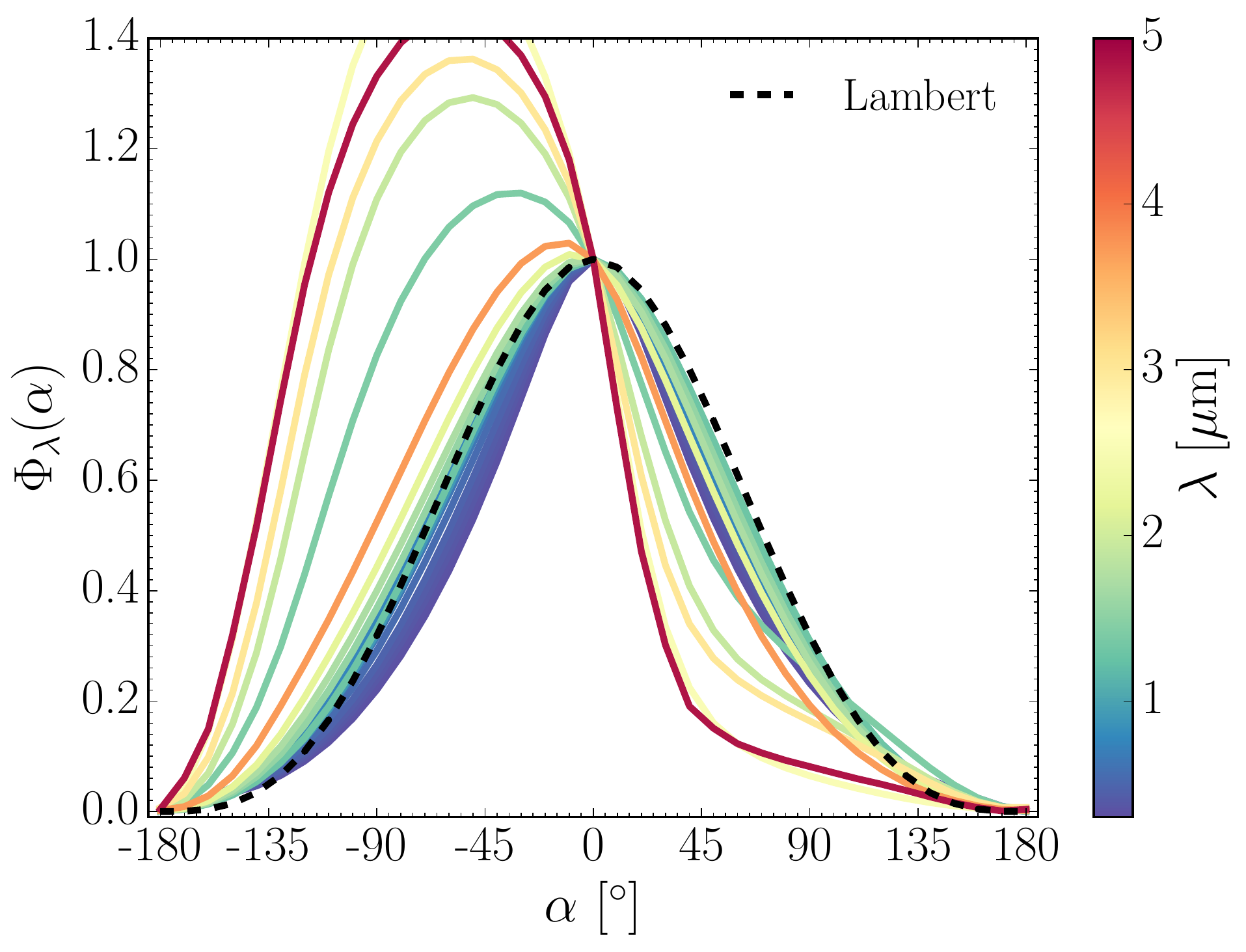} 
   \caption{Left: Scattered light phase curves of the albedo spectra A$_{\lambda}$($\alpha$) (Eq. \ref{eq:albspec}) of our HD 189733b simulation between 0.3 $\mu$m (colour bar: dark purple) to 5.0 $\mu$m (colour bar: dark red).
   Optical and near-IR wavelength packets are more strongly scattered than longer IR wavelengths.
   Right: Classical phase function (Eq. \ref{eq:phasefunction}), which emphasises the phase curve shapes. 
   Optical wavelengths are more symmetric about the sub-stellar point, IR wavelengths show more variation with phase.
   A Lambertian phase function (black, dashed line) is over-plotted for reference, which is useful for characterising scattering behaviour (see text).}
   \label{fig:Ophase}
\end{figure*}

Figure \ref{fig:Ophase} (left) presents the albedo A$_{\lambda}$($\alpha$) (Eq. \ref{eq:albspec}) phase curves at 25 different wavelengths in the range considered in the MCRT post-processing.
Optical and some near-IR wavelengths show scattering fractions A$_{\lambda}$($\alpha$) $>$ 0.1 across $-$90\degr$\ldots$90\degr\ viewing angles.
This suggests that a significant percentage of the planets optical scattered luminosity remains observable, while a majority of the dayside hemisphere remains in view.
Scattered light fractions drop below 10\% for $\lambda$ $>$ 3 $\mu$m.

Figure \ref{fig:Ophase} (right) shows the classical scattered light phase function $\Phi_{\lambda}$($\alpha$) (Eq. \ref{eq:phasefunction}).
This adds emphasis to any asymmetric scattering behaviour as a function of phase.
Optical and near-IR wavelengths 0.3-1$\mu$m show very symmetric scattering phase curves around the sub-stellar point, while longer 1-5$\mu$m can show asymmetry in their scattering properties.
Longer IR wavelengths can also show westward offsets of 10-20\degr, depending on wavelength.
This is due to larger cloud particles residing on the western hemisphere of the dayside (Paper I), increasing the opacity and scattering probability of IR packets travelling in these regions.

We compare our classical phase function to the Lambertian phase function (Fig. \ref{fig:Ophase}, dashed black line), which is the phase function of a theoretical, perfectly isotropic scattering sphere \citep[e.g.][]{Seager2010, Madhusudhan2012}.
This is useful for interpreting the type of scattering behaviour at east and west dayside hemispheres \citep{Madhusudhan2012}.
The eastern hemisphere shows a classical phase function behaviour typical of Rayleigh-like scattering across all wavelengths, suggesting that the majority of optical and near-IR packets undergo a single scattering event in these regions.
This region contains the smallest cloud particle sizes in the RHD simulation and lowest cloud opacity, increasing the likelihood of the packet escaping the boundaries without further interaction.
The western hemisphere shows a mix of isotropic and Rayleigh-like scattering classical phase function behaviour for optical, near-IR and IR wavelengths.
This suggests that packets experience multiple scattering events, which results in packets escaping in an isotropic manner.
These regions contain some of the larger cloud particles on the dayside and an increased cloud scattering probability.

Figure \ref{fig:Ophase} (right) also shows IR (3-5 $\mu$m) classical phase function asymmetry that arises from the differences in the opacity structure between the two day-night terminator regions.
The westward limb ($\alpha$ = $-$90\degr) contains larger cloud particle sizes, opacity, and scattering probability in the IR compared to the eastern limb ($\alpha$ = 90\degr).
The behaviour of the IR wavelength classical phase functions suggests that packets do not interact with the eastern terminator cloud structures and pass through without interaction, are absorbed, or interact with cloud particles on the nightside hemisphere since there is a return to the Lambertian function at nightside viewing angles.
For the westward limb, the IR cloud opacity is higher, packets travel shallower into the atmosphere before interacting and are more likely to be scattered.
These differences in cloud structures between the terminator regions result in a skewing of the IR classical phase function.
The western side of the planet is therefore typically brighter in the IR by 10-20\% in scattered light than the eastern side at comparable phases.

\subsection{Emitted light spectra and phase curves}
\label{sec:emitted}

\begin{figure*}
   \centering
   \includegraphics[width=0.75\textwidth]{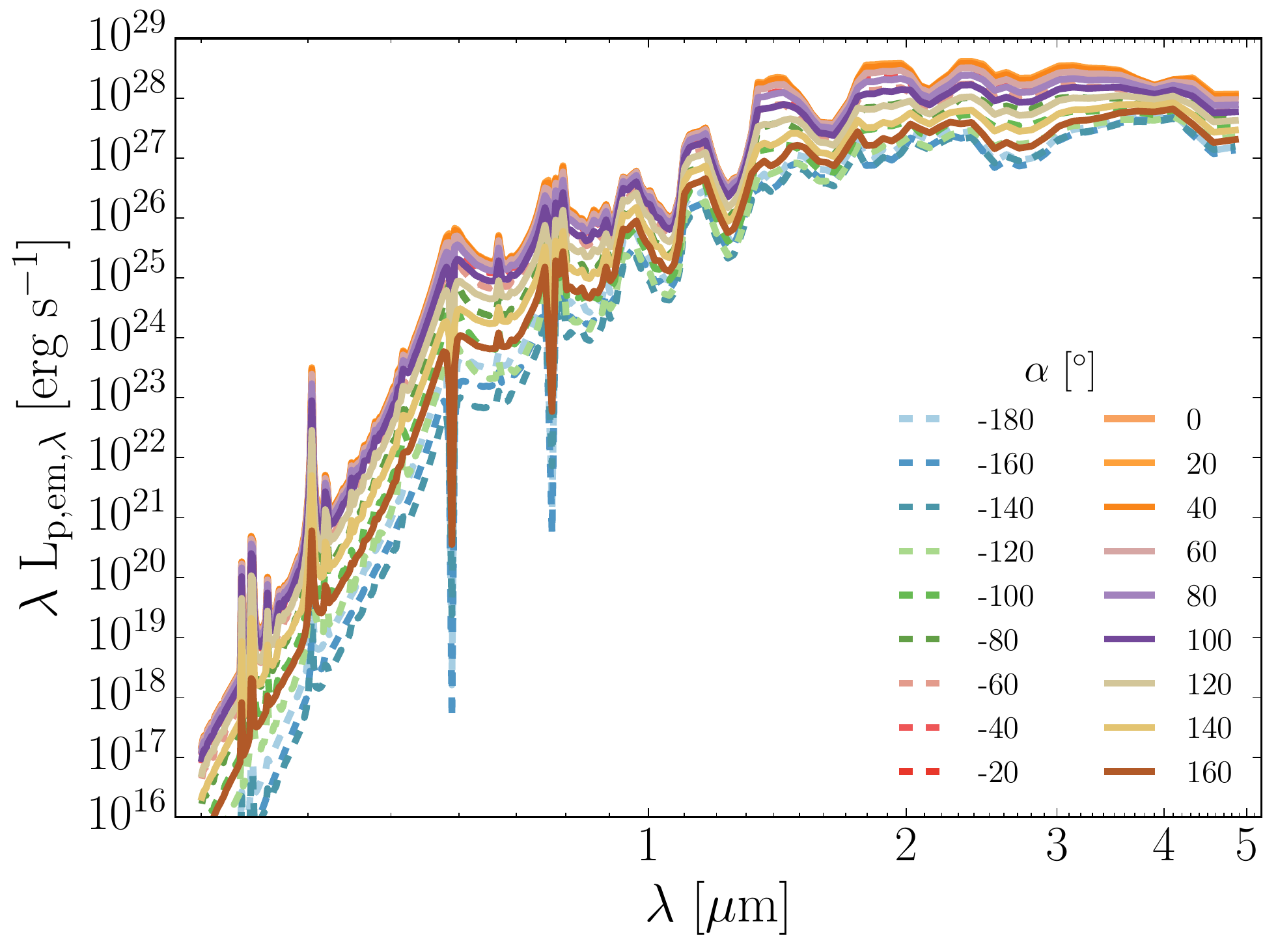} 
   \caption{Emitted light spectral energy distribution $\lambda$ L$_{\rm p,em,\lambda}$ (Eq. \ref{eq:emitted}) of our HD 189733b simulation as a function of wavelength at several viewing angles $\alpha$ ($\alpha$ = 0\degr\ is the sub-stellar point).
   Spectral features remain qualitatively similar at different viewing angles with a $\sim$ 1 order of magnitude difference between the sub-stellar point ($\alpha$ = 0\degr) and anti-stellar point ($\alpha$ = $-$180\degr).
   The peak wavelength is found to be at $\sim$ 2 $\mu$m.}
   \label{fig:Tphase_spec}
\end{figure*}

\begin{figure*}
   \centering
   \includegraphics[width=0.49\textwidth]{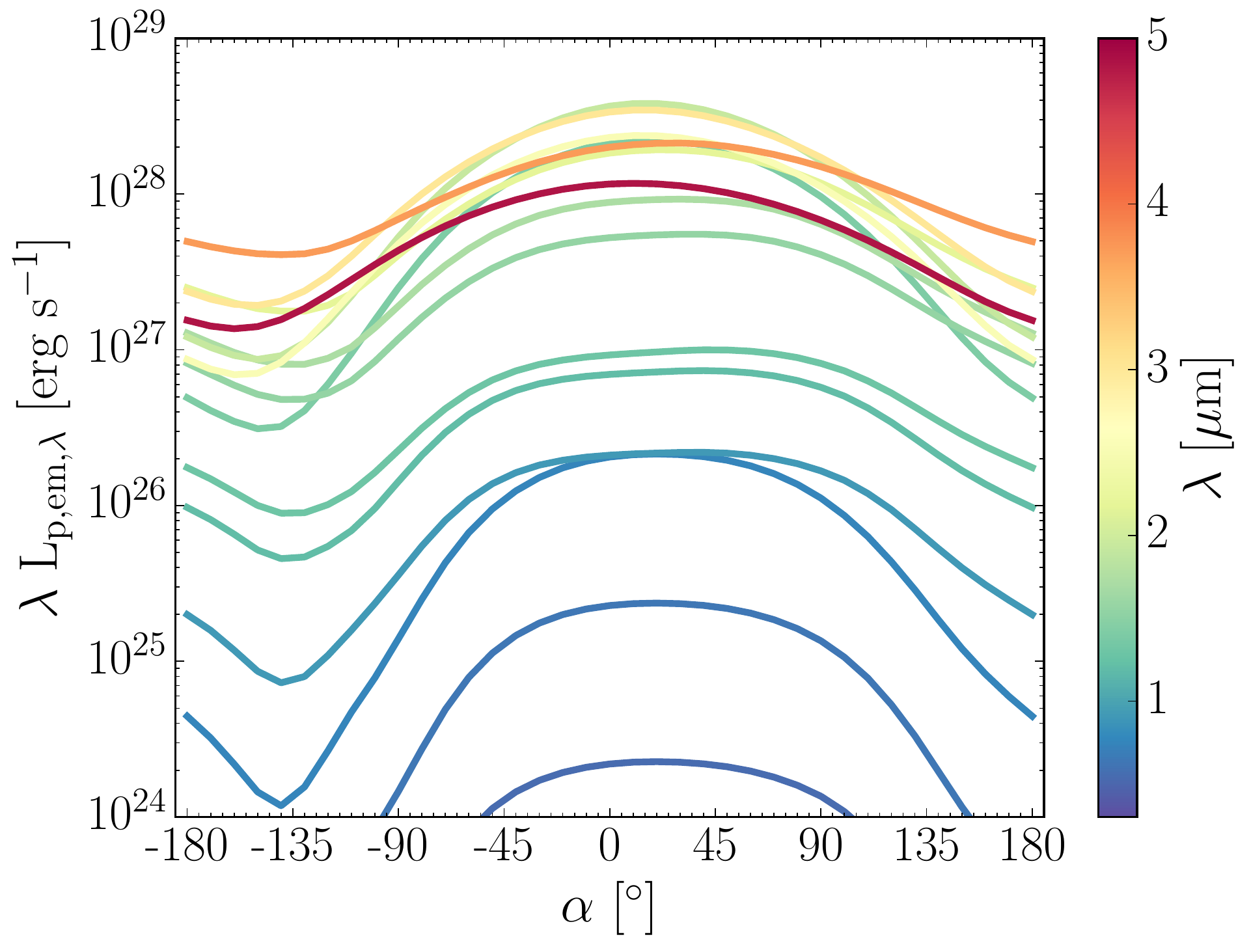}       
   \includegraphics[width=0.49\textwidth]{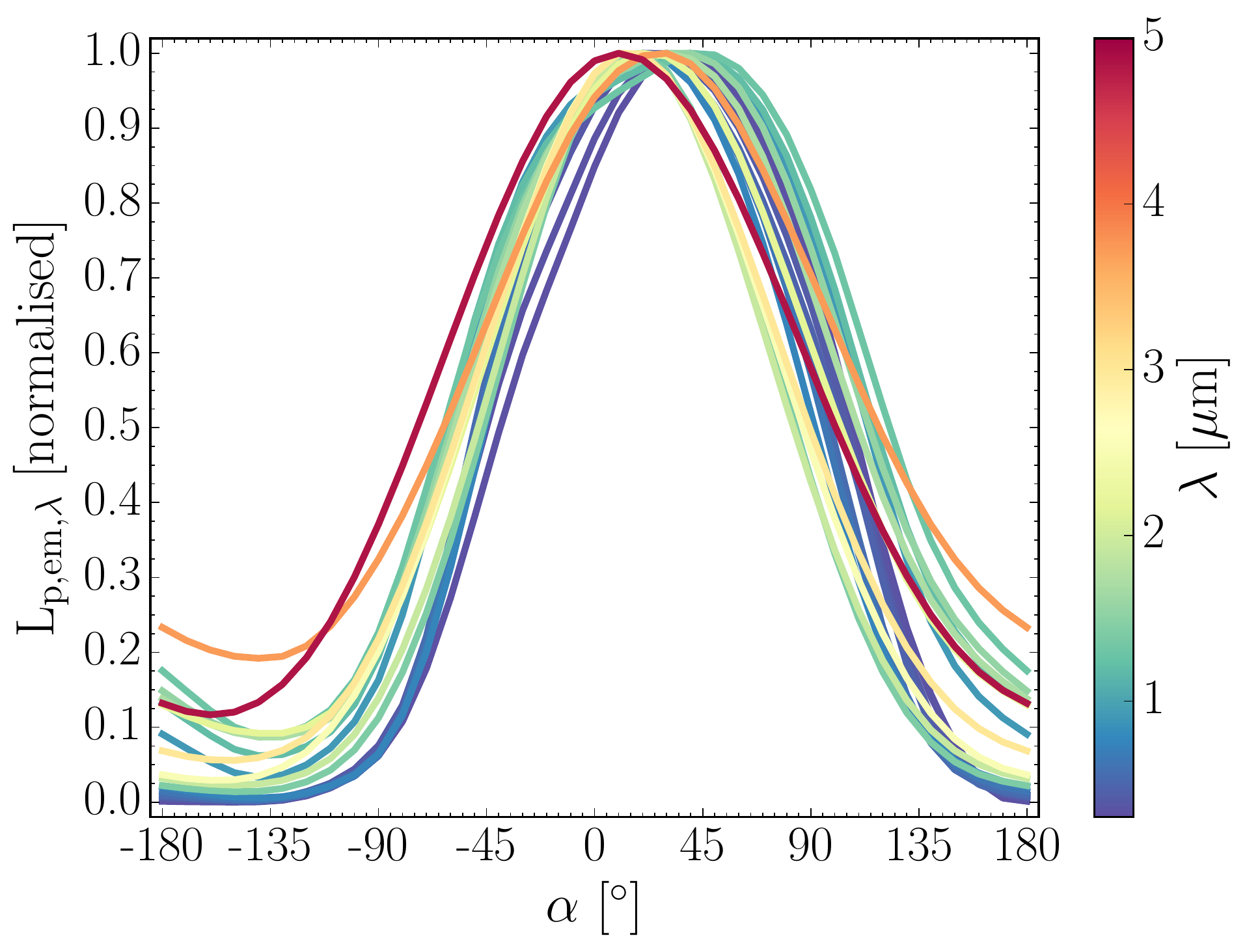} 
   \caption{Left: Emitted light spectral energy distribution $\lambda$ L$_{\rm p,em,\lambda}$ (Eq. \ref{eq:emitted}) of our HD 189733b simulation phase curves between 0.3 $\mu$m (colour bar: dark purple) to 5.0 $\mu$m (colour bar: dark red).
   Infrared wavelengths from 3-5 $\mu$m dominate the emission luminosity at all phases.
   Right: Normalised (to maximum $\lambda$ L$_{p,em,\lambda}$) spectral energy distribution phase curves to emphasise the phase curve shapes.
   All wavelengths show a $\alpha$ $\ge$ 10\degr\ eastward offset from the sub-stellar point.}
   \label{fig:Tphase}
\end{figure*}

\begin{figure*}
   \centering
   \includegraphics[width=0.75\textwidth]{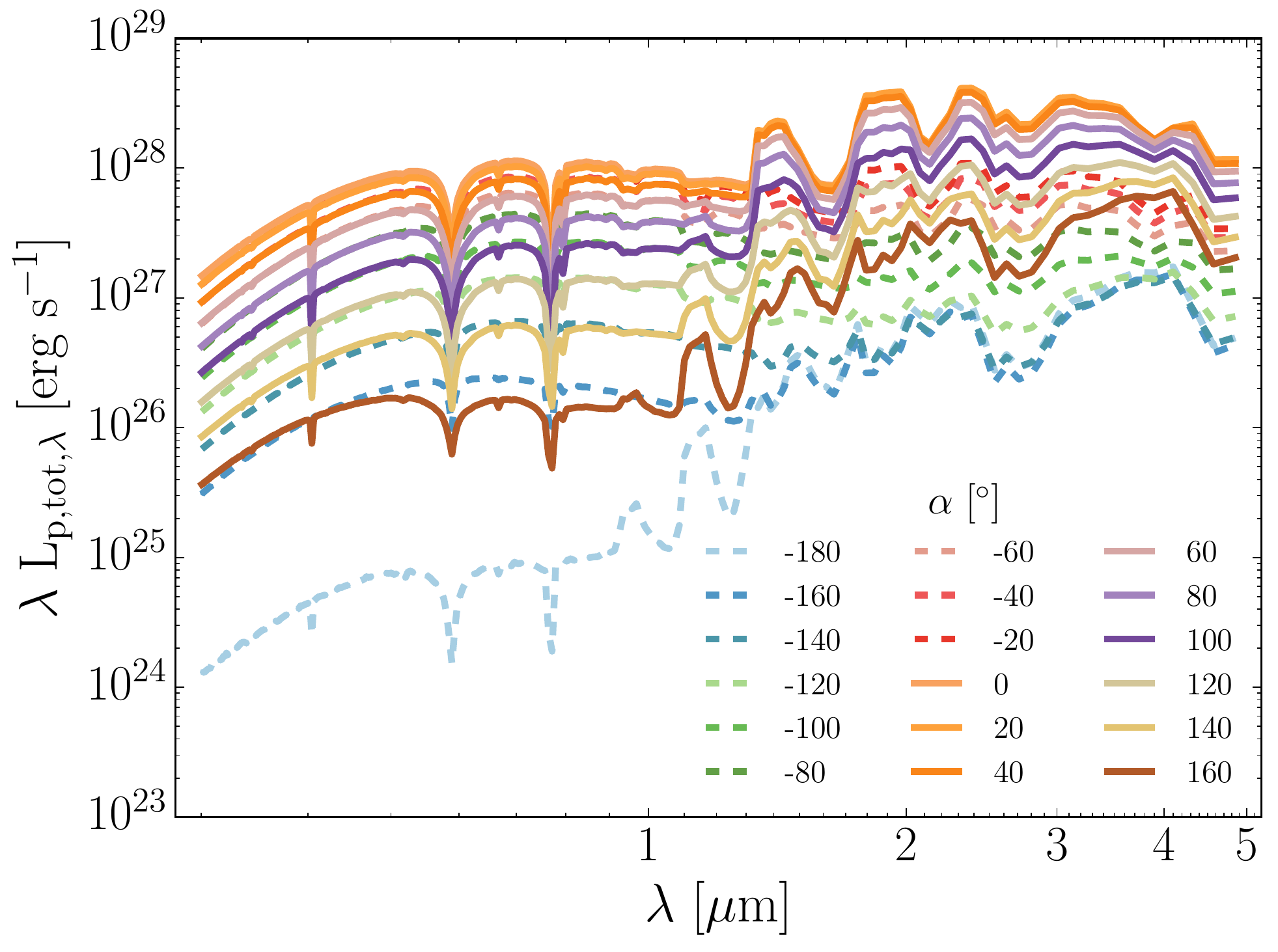} 
   \caption{Combined scattered and emitted light spectral energy distribution $\lambda$ L$_{\rm p,tot,\lambda}$ (Eq. \ref{sec:combined}) of our HD 189733b simulation as a function of wavelength at several viewing angles $\alpha$ ($\alpha$ = 0\degr is the sub-stellar point).
   The optical wavelength scattered light luminosity drops off as the nightside of the planet is viewed.
   The emission features between the dayside and nightside are $\sim$ 1 order of magnitude different dependent on wavelength.}
   \label{fig:comb_phase_spec}
\end{figure*}

At secondary transit, observations are also sensitive to the thermal emission from the planetary atmosphere itself.
Because the star-planet contrast is larger at IR wavelengths than in the optical, the planet's atmospheric emission features are easier to detect.

Figure \ref{fig:Tphase_spec} shows the spectral energy distribution (SED) $\lambda$ L$_{\rm p, em \lambda}$ [erg s$^{-1}$] (Eq. \ref{eq:emitted}) as a function of wavelength for the atmospheric emitted light.
We find that the SED follows similar trends across all viewing angles with $\sim$ 1$-$2 magnitude differences between dayside and nightside emission luminosities.
The peak of emission occurs at $\sim$ 2 $\mu$m, where our model HD 189733b is brightest in emitted light.
The least emission luminosity in the IR occurs at a viewing angle of $\alpha$ = $-$140\degr, corresponding to the coldest regions of the nightside.

Figure \ref{fig:Tphase} (left) presents the planetary atmosphere SED $\lambda$ L$_{\rm p, em, \lambda}$ (Eq. \ref{eq:emitted}) as a function of viewing angle $\alpha$.
We find that emission is dominated by the longer IR wavelengths 2-5$\mu$m for all phases.
Emission at optical wavelengths is confined to the hottest dayside regions of the planet, however, this emission is $>$ 4 magnitudes in luminosity smaller than IR wavelengths.
A dip in luminosity is seen at $\sim$ $-$135\degr for all wavelengths, corresponding to the coldest nightside regions of the RHD simulation.
Figure \ref{fig:Tphase} (right) shows the normalised emitted light phase curves, which emphasises the shape of the phase curve.
Most wavelengths show eastern offsets from the sub-stellar point in peak emission at 10-25\degr with some IR wavelengths showing offsets $\sim$ 45\degr from the sub-stellar point.

We find that the effects of atmospheric scattering by emitted packets on the phase curves to be negligible.
This is because of the low gas phase single scattering albedos ($\omega_{\rm gas}$ $<$ 10$^{-3}$) at IR wavelengths, which lowers the weights of the packets to negligible proportions after one or two scattering interactions.
The majority of packets are then likely to be terminated by the Russian Roulette scheme, and those that survive will contain low luminosity weights in future interactions.

\subsection{Total luminosities and observational predictions}
\label{sec:res_combine}

\begin{figure*}
   \centering
   \includegraphics[width=0.75\textwidth]{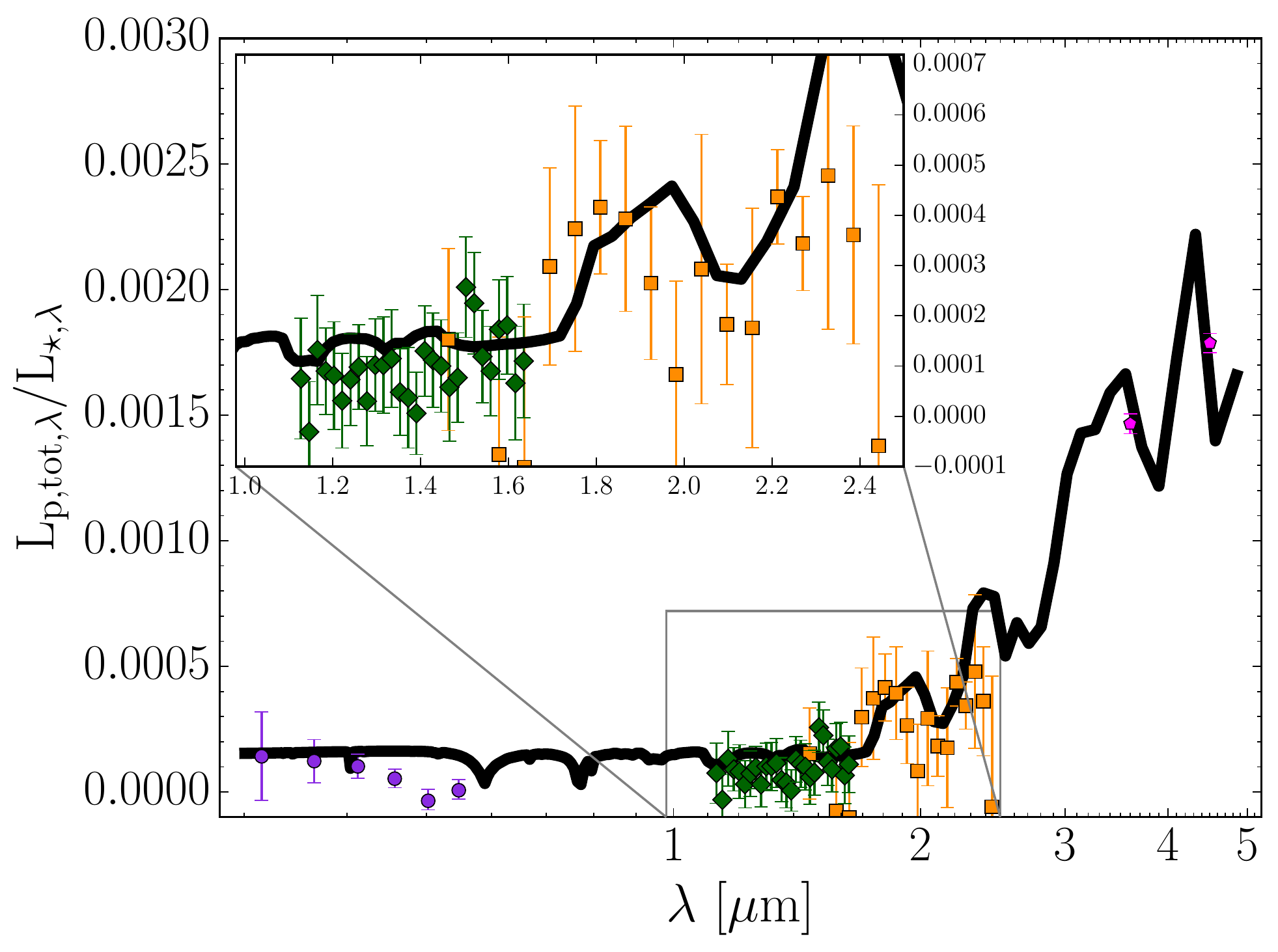} 
   \caption{Combined scattered and emitted light flux ratio L$_{\rm p,tot,\lambda}$/L$_{\star.\lambda}$ of our HD 189733b simulation as a function of wavelength at secondary eclipse ($\alpha$ = 0\degr).
   We compare our HD 189733b simulation results to available observational data from HST STIS (purple dots, 0.29-0.57 $\mu$m) \citep{Evans2013}, HST WFC3 (green diamonds, 1.13-1.64 $\mu$m) \citep{Crouzet2014}, HST NICMOS (orange squares, 1.46-2.44 $\mu$m) \citep{Swain2009, Barstow2014}, and Spitzer IRAC (magenta pentagons, 3.6 $\mu$m, 4.5 $\mu$m) \citep{Knutson2012}. }
   \label{fig:comb_phase_ratio}
\end{figure*}

\begin{figure*}
   \centering
   \includegraphics[width=0.49\textwidth]{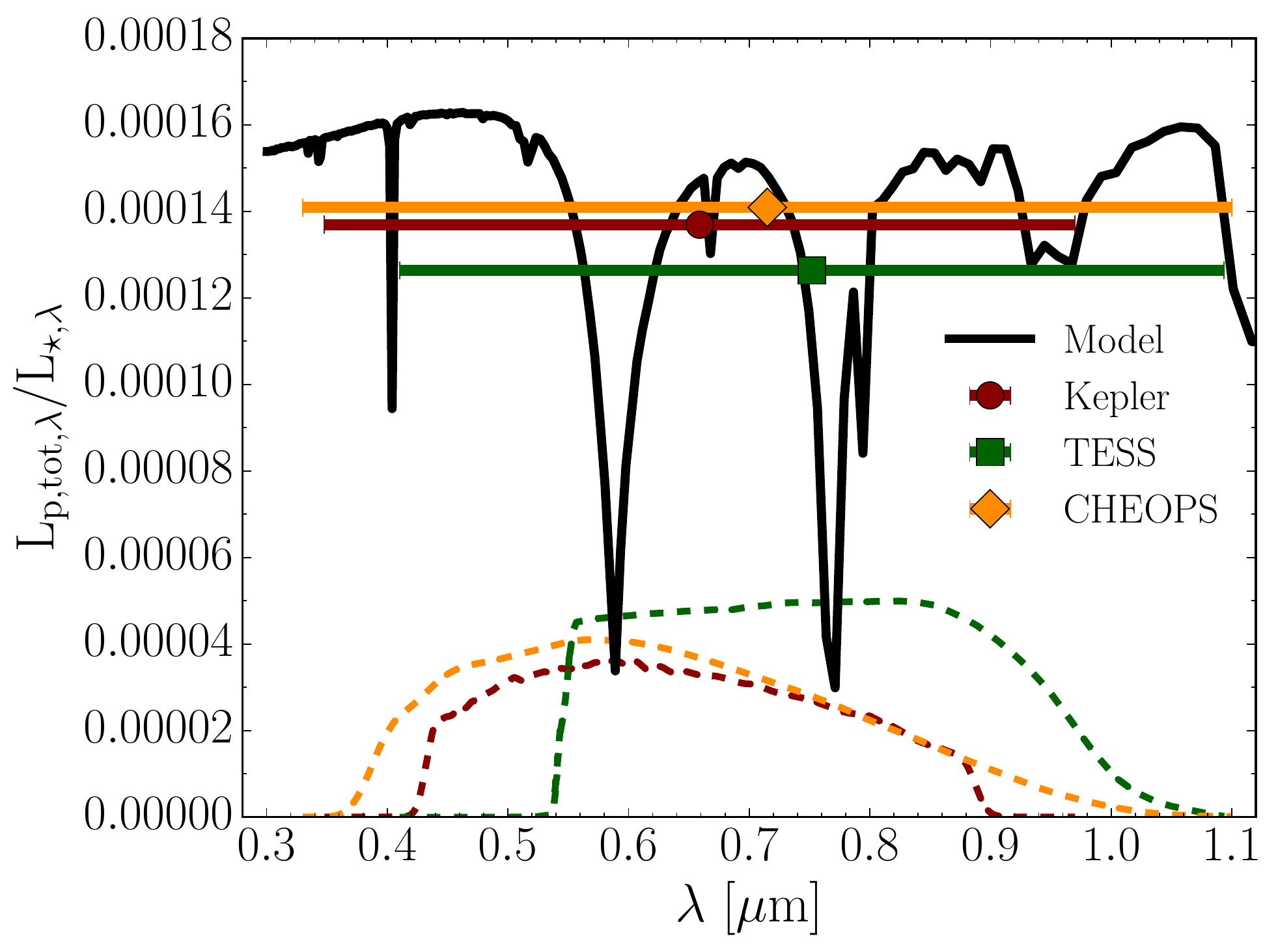} 
    \includegraphics[width=0.49\textwidth]{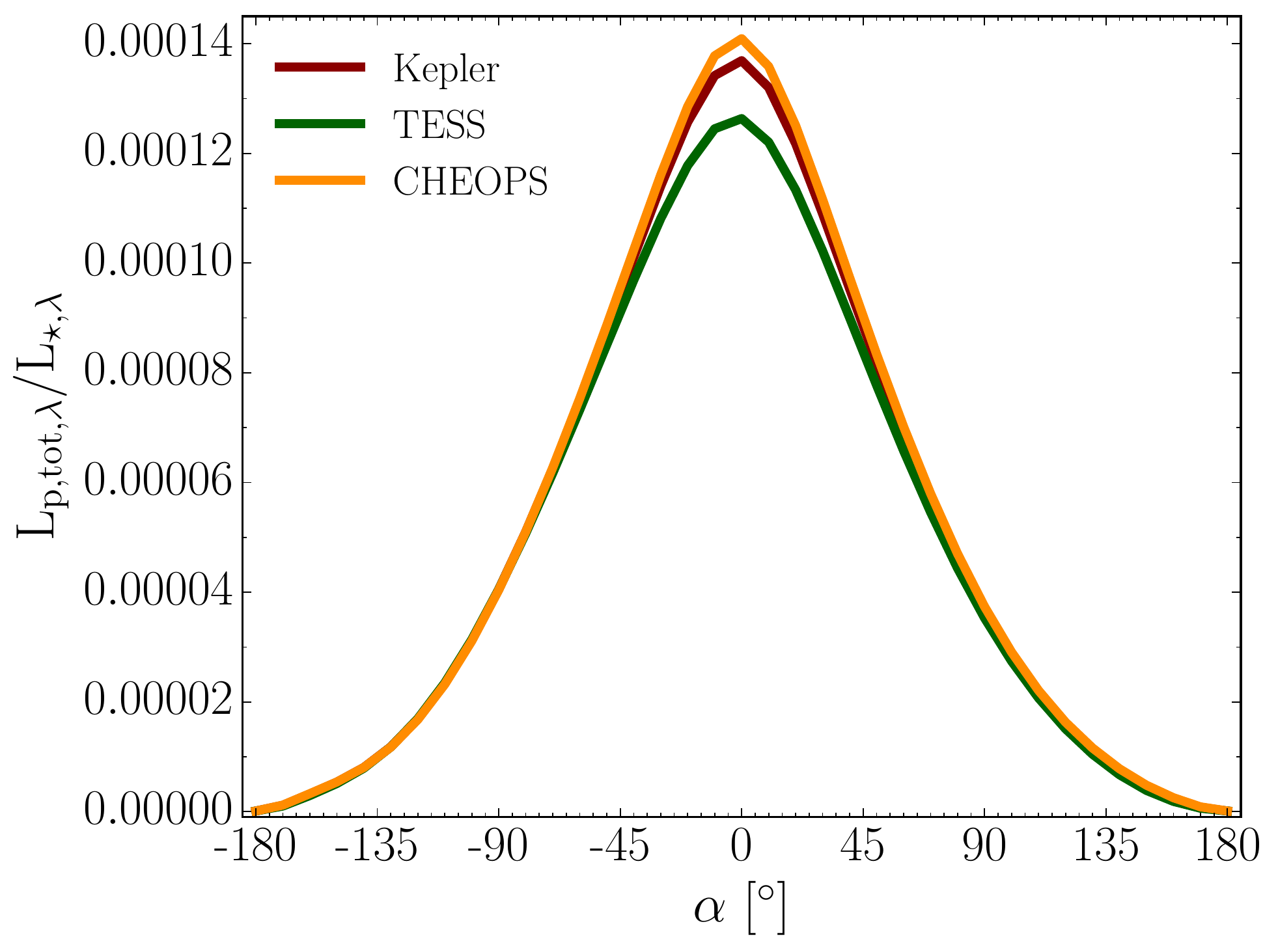} 
   \caption{Combined scattered and emitted light flux ratio L$_{\rm p,tot,\lambda}$/L$_{\star.\lambda}$ HD 189733b predictions for the Kepler, TESS, and CHEOPS instruments.
   Left: Predicted dayside flux ratio for Kepler (red dot), TESS (green square) and CHEOPS (orange diamond) photometric bandpasses. The solid black line is the model output. Dashed lines indicate the spectral response function of the instrument.
   Right: Predicted Kepler (red), TESS (green) and CHEOPS (orange) flux ratio phase curves. 
   All three instruments show no/little offset from the sub-stellar point, we predict a westward maximum flux offset of only -5\degr$-$ 0\degr from the sub-stellar point.
    }
   \label{fig:comb_phase_pred}
\end{figure*}

The observed luminosity from an exoplanet atmosphere is a combination of the reflected starlight and the emission from the planet itself.
The component that dominates the observation will be wavelength- and phase-dependent owing to the 3D inhomogeneous scattering opacities (from cloud particles) and temperature profiles (from atmospheric circulation) of the planet.
Such inhomogeneity is present in the results of the HD 189733b RHD cloud forming simulation.
In this section, we combine our scattering and thermal emission results above to produce observable quantities in the TESS \citep{Ricker2014} and CHEOPS \citep{Broeg2013} bandpasses for our HD 189733b model.
We also compare to readily available observational data from current/past missions.
We also produce Kepler band predictions, despite K2 not scheduled to observe HD 189733b to allow a comparison to other studies focused on modelling Kepler objects \citep[e.g.][]{Parmentier2016}.

The total monochromatic luminosity from the planet is the sum of the reflected light and emitted light
\begin{equation}
\label{sec:combined}
L_{\rm p, tot, \lambda} = L_{\rm p, scat, \lambda} + L_{\rm p, em, \lambda}.
\end{equation}
The apparent geometric albedo (which includes scattered and emitted light) is then
\begin{equation}
\label{eq:comb_ag}
A_{g, \lambda} = \frac{L_{\rm p, tot, \lambda}(\alpha = 0\degr)}{L_{\rm inc, \lambda} 2/3} ,
\end{equation}
in which the numerator and denominator are integrated across the Kepler, TESS, and CHEOPS bandpasses.
For our comparisons to observational data and observational predictions, instead of assuming the host star radiates as a black body in previous sections, the stellar monochromatic luminosity of HD 189733 (L$_{\star, \lambda}$) is taken from the Kurucz \footnote{http://kurucz.harvard.edu/stars/hd189733/} stellar atmosphere model for HD 189733.

Figure \ref{fig:comb_phase_spec} shows the spectral luminosity of the planet as a function of phase including both the scattering and emitted components.
On the dayside of the planet, the luminosity from scattered light is at a similar magnitude to the IR emission.
For nightside profiles, the scattered light fraction drops off and the emitted luminosity dominates.

Figure \ref{fig:comb_phase_ratio} presents the flux ratio L$_{\rm p, tot, \lambda}$/L$_{\star, \lambda}$ of our combined scattered and emitted light, compared to the HST secondary eclipse data from \citet{Swain2009, Evans2013,Crouzet2014,Barstow2014} and Spitzer data from \citet{Knutson2012}.
Similar to Sect. \ref{sec:geoalb}, our simulation is consistent with the B Band HST STIS data from \citet{Evans2013} but overestimates the planet-star flux ratio for the V Band.
The results are consistent with the HST WFC3 \citep{Crouzet2014} and NICMOS \citep{Swain2009, Barstow2014} spectral trends, with offsets in the amplitude of these features. 
Several retrieval models on this and similar observational data \citep[e.g.][]{Madhusudhan2009,Lee2012,Line2012,Line2014} suggest a possible sub-solar H$_{2}$O abundance on the dayside of HD 189733b, which would lower the amplitude of the H$_{2}$O features in our results.
Our results are also consistent with the 3.6 $\mu$m and 4.5 $\mu$m Spitzer IRAC photometry observations of \citet{Knutson2012}. 
Overall, our current model reproduces the secondary transit optical to near-IR observational trends well.

Figure \ref{fig:comb_phase_pred} (left) shows the dayside flux ratio predictions for Kepler, TESS, and CHEOPS bandpasses.
We predict a $\sim$ 10 \% difference in peak flux ratio between the CHEOPS and TESS bandpasses.
This is directly due to the sensitivity of the CHEOPS bandpass to the optical scattering component of our RHD modelled cloud particles, while the TESS bandpass is unaffected by this component and more sensitive to the near-IR thermal emission of the HD 189733b model.

Figure \ref{fig:comb_phase_pred} (right) presents the flux ratio phase curves for the Kepler, TESS, and CHEOPS bandpasses.
Our modelling results show that we expect HD 189733b to have a zero or small westward offset of no less than -5\degr\ from the sub-stellar point for the Kepler, TESS, and CHEOPS bands.
This suggests that the cloud particle differences (size, composition, etc.) at the east and western hemispheres in our RHD simulation are not radically different enough to produce the larger ($<$ $-$10\degr) westward offsets seen for some Kepler planets.
The TESS and CHEOPS photometry become comparable at greater longitudes as the cloud particle scattering component becomes less dominant compared to the thermal emission (Fig. \ref{fig:comb_phase_spec}).

If the extra absorption component from the \citet{Evans2013} HST STIS measurements at $\sim$ 0.4-0.5 $\mu$m is taken into account, owing to the bandpass efficiencies, the Kepler and CHEOPS contrast ratios and geometric albedos of HD 189733b are likely to be lower than those presented here.
However, the TESS photometric band would be relatively unaffected owing to the low sensitivity in this wavelength regime, unless the influence of this absorber extends into the near-IR.
Table \ref{tab:sum} summarises our Kepler, TESS, and CHEOPS predictions, including estimates of the geometric albedo A$_{g}$ (Eq. \ref{eq:comb_ag}) in each respective band.

\begin{table}[htdp]
\caption{Summary of observational predictions. A$_{g}$ is the predicted geometric albedo across the bandpass. The offset is defined as degrees from the sub-stellar point. }
\begin{center}
\begin{tabular}{|c|c|c|c|}
Telescope & $\lambda$ [$\mu$m] & A$_{g}$ & Offset [\degr] \\ \hline
Kepler & 0.35-0.97 & 0.2221 & $-$5$-$0 \\ 
TESS & 0.41-1.10 & 0.2050 & $-$5$-$0 \\
CHEOPS & 0.33-1.10 & 0.2286 & $-$5$-$0
\end{tabular}
\end{center}
\label{tab:sum}
\end{table}%

\subsection{Variance and convergence}
\label{sec:var}

\begin{figure}
   \centering
   \includegraphics[width=0.49\textwidth]{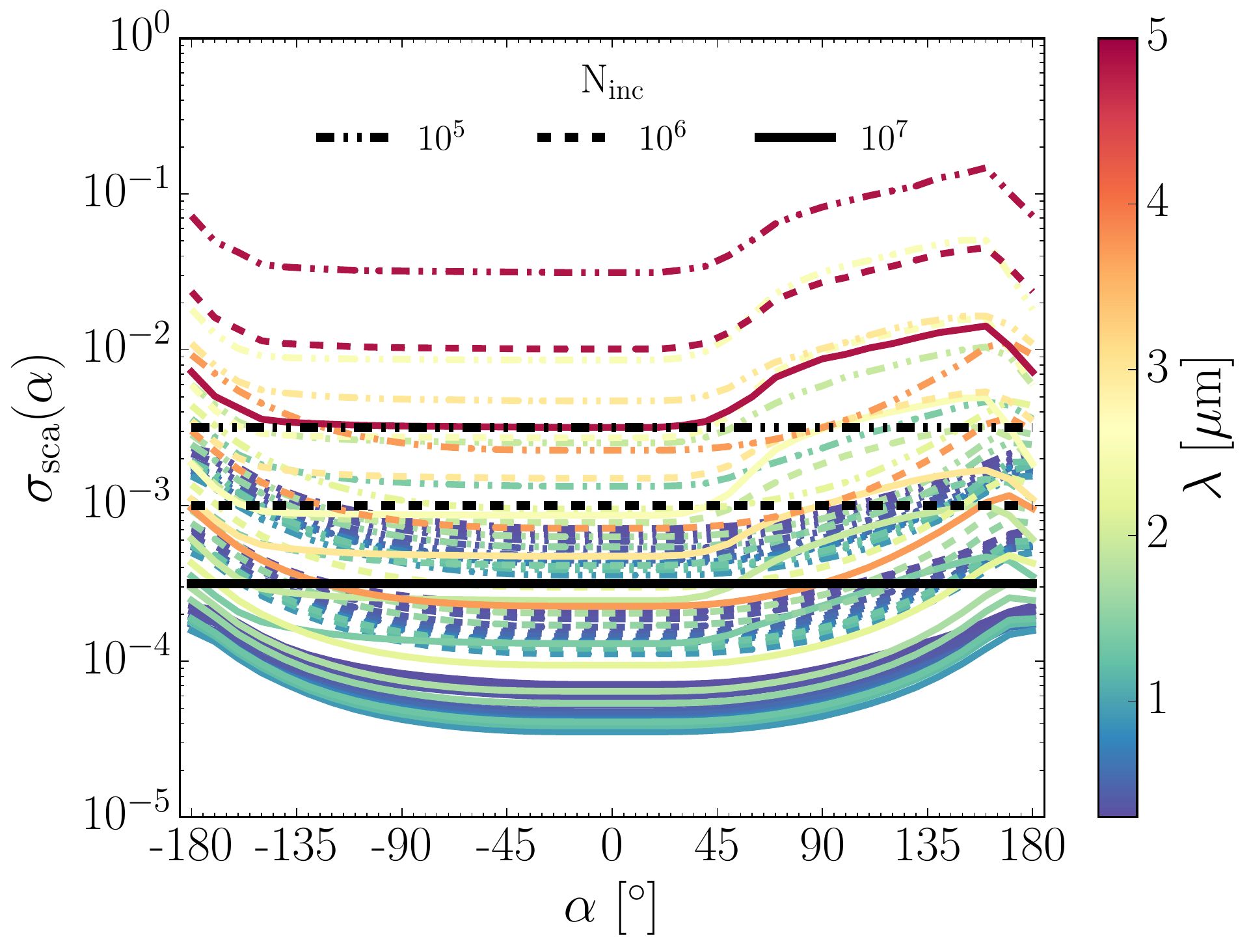} 
   \includegraphics[width=0.49\textwidth]{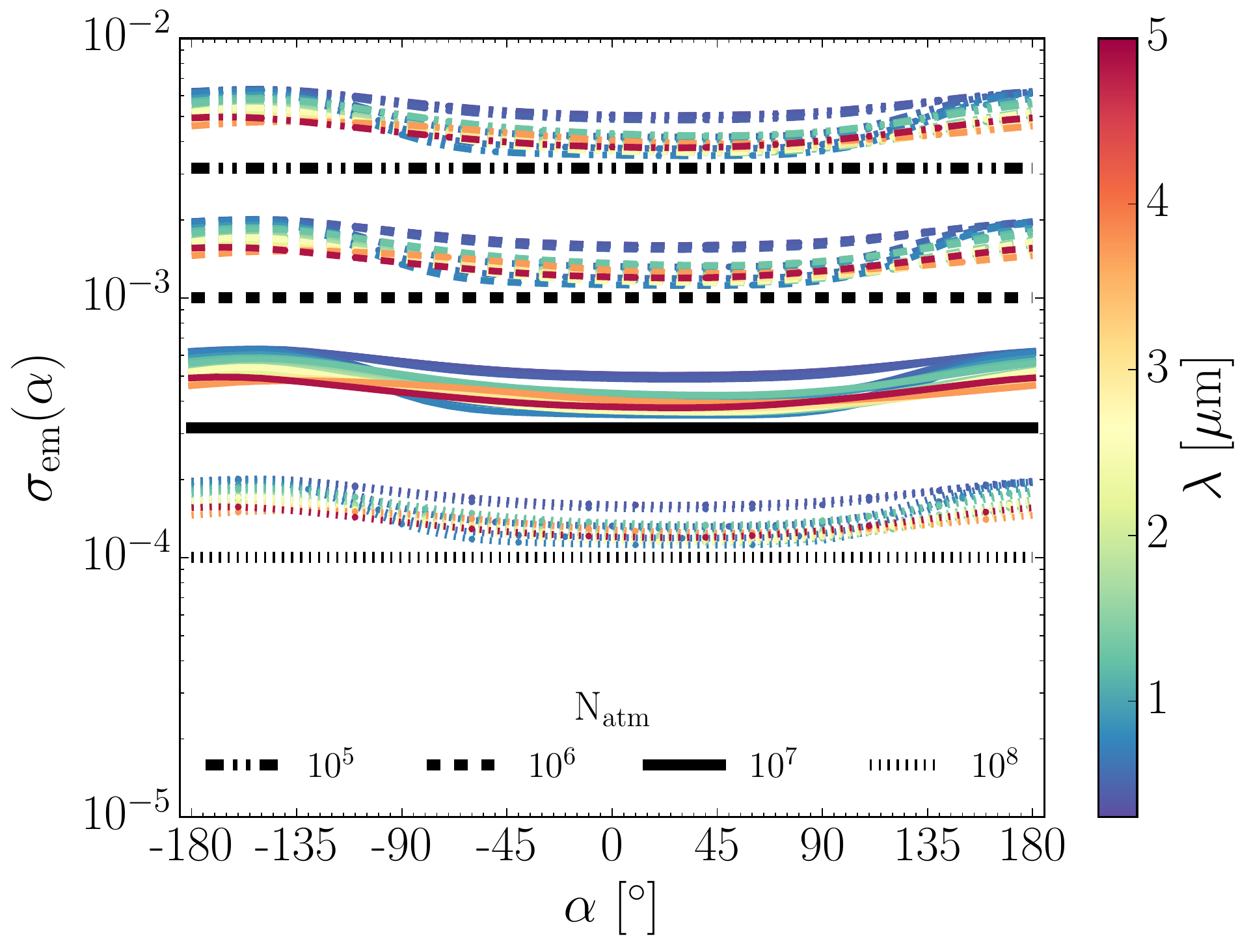} 
   \caption{Scattered light (Top) and emitted light (Bottom), noise error $\sigma$ (Eq. \ref{eq:var}) with viewing angle $\alpha$ and wavelength (colour bar).
   The nominal noise error values for $N_{inc/atm}$ = 10$^{5}$ (dash-dot line), 10$^{6}$ (dashed line), 10$^{7}$ (solid line) and 10$^{8}$ (dotted line) are the horizontal black lines.
   Most scattered and emitted light images are near or below the nominal values, suggesting that each packet contributes once or more to multiple images.}
   \label{fig:var}
\end{figure}

We conclude the results section with a discussion on the noise errors and convergence of the MCRT results. 
Because of the stochastic nature of MCRT, it exhibits Poisson sampling statistics, where the noise error $\sigma$ is given by 
\begin{equation}
\sigma = \frac{1}{\sqrt{N_{\rm im}}} ,
\label{eq:var}
\end{equation}
where $N_{\rm im}$($\lambda$, $\alpha$) is the number of photon packets contributing to the output image, dependent on the wavelength and viewing angle.
To test the convergence of the results, the same experiments are performed with a varying number of initialised photon packets. 
We repeat the above simulations with $N_{\rm inc/atm}$ = 10$^{5}$, 10$^{6}$ and 10$^{7}$  to test the decrease of $\sigma$ with initialised packet numbers.

Figure \ref{fig:var} shows the noise error of each scattered and emitted light image for our three $N_{\rm inc/atm}$ simulations with viewing angle $\alpha$ and wavelength.
We also show the nominal value (black horizontal line) from the total initialised photons ($N_{\rm inc}$, $N_{\rm atm}$). 
In incident scattered light, all but the longest IR wavelengths show noise errors below the nominal values, indicating that most incident stellar packets contribute to one or more image results.
In thermal emitted light, the noise error values are within 2-3 times the nominal value, suggesting that most thermally emitted photons also contribute to many image outputs.
The error only increases by a factor of 2 or 3 for images including a nightside component, which suggests that composite emission biasing was successful in more evenly spreading the noise error between high and low luminosity regions.
Our presented results of the previous sections using $N_{\rm inc}$ = 10$^{7}$, $N_{\rm atm}$ = 10$^{8}$, have a typical model noise error of $<$ 0.1 $\%$.

\section{Discussion}
\label{sec:Discussion}

In this section we discuss the overall context of our results both theoretically and observationally, and detail some of the limitations of the current simulations.
In summary, our updated model improves the realism of the \citet{Hood2008} scheme in several ways:

\begin{itemize}
\item Cloud particle opacity and scattering properties, derived from 3D self-consistent cloud forming RHD simulation results.
\item Addition of a thermal emission mode, using the thermodynamic properties of the 3D RHD simulation results.
\item Higher radial, longitude, latitude resolution (R, $\phi$, $\theta$) = (100, 160, 64) given by the RHD grid cell sizes.
\item k-distribution opaque absorbing gas species to the opacity scheme interpolated to the 3D RHD T-p profiles.
\item Addition of survival biasing and composite biasing variance reduction techniques for the MCRT scheme.
\item Stochastic cloud particle size distribution sampling, based on the microphysical cloud formation results of the RHD simulation.
\item Rayleigh scattering for small cloud particle size parameters and H$_{2}$ molecular scattering.
\item Two-term Henyey-Greenstein scattering for larger size parameter cloud particle scattering events.
\end{itemize}

In \citet{Hood2008} a fractal scheme was introduced to their simplified cloudy atmosphere structure to model "patchiness" in small scale cloud properties.
These authors found that increasing the porosity of their original smoothly distributed cloud structure decreased the geometric albedo results, as L-packets passed more easily through cloudy regions to the deeper atmosphere.
Applying a similar technique to the current study may reconcile the differences in geometric albedo of the model to the observational data.
We note that the amount of porosity is not known a priori, however.

Our current study depends on the 3D cloud forming HD 189733b RHD model results of \citet{Lee2016}.
We summarise the limitations of that study and the possible effects on the MCRT results presented here.
The long-term (1000s simulated days) settling timescales of the sub-micron cloud particles in the upper atmosphere is not captured by our $\sim$ 60 Earth days of total integration \citep[e.g.][]{Woitke2003,Parmentier2013}.
The possible removal of sub-micron grains from the upper atmosphere would have two main effects:
lowering the overall opacity of the upper atmospheric cloud and increasing the likelihood of cloud particle Rayleigh scattering events as the size distribution is further skewed towards smaller particle sizes.
However, sub-micron grains are likely to remain lofted over long timescale in the atmospheres of hot Jupiters from hydrodynamical motions alone \citep{Parmentier2013}, suggesting that a total rain out of sub-micron cloud particles is an unlikely scenario.

Our gas phase opacities in Paper I used Planck mean opacities, which have been shown to have band averaged errors for the stellar heating rates \citep{Amundsen2014}. 
This may be offset by the greyer opacity of the cloud particles, but regions where the gas opacity dominates (e.g. high-temperature dayside regions) these errors may remain.
Although in this post-processing study we use k-distribution tables, the cloud properties derived from the RHD model may be affected by this choice of opacity scheme.
We suggest in \citet{Lee2016} that the use of the correlated-k approximation would alter the depth of the seed particle regions on the dayside of the RHD model, dependent on whether the dayside atmosphere is cooler or warmer because of this opacity change.
This would alter the cloud opacity structure; the effect on the MCRT would most likely be an increased or decreased luminosity escaping from the next-event estimator scheme, depending on whether the main scattering regions were higher or lower in atmospheric optical depth. 
Additionally, the \citet{Lee2016} simulation assumed an absorptive opacity, neglecting scattering effects of clouds particles, which would also impact the thermal structure of the RHD results.

Fe[s] materials were also not modelled in \citet{Lee2016}, which may also increase the opacity of cloud particles and reduce the single-scattering albedo of cloud particles owing to the absorbing properties of Fe[s] materials.
However, Fe[s] rich grains are expected to reside below the photosphere of HD 189733b \citep{Lee2015b, Helling2016}, where they would have negligible effects on the observable scattered and emitted light.

Our model does not reproduce the low V Band geometric albedo observed by \citet{Evans2013}.
This low geometric albedo has been noted by other studies \citep[e.g.][]{Barstow2014}, with a suggestion that another, unknown absorber is responsible for the sharp decrease in albedo at 0.5 $\mu$m.
\citet{Barstow2014} modelled several scenarios that could be responsible for this decrease: super-solar Na abundance, which would extend the Na absorption wings; TiO absorption; and also MnS[s] cloud compositions.
If the effects of this unknown absorber extend into the near-IR wavelength regime, we can expect a lower geometric albedo for the TESS and CHEOPS bandpasses than the chemical equilibrium, solar-metallicity gas opacity calculations applied here.

Our CHEOPS and TESS predictions show that we predict the CHEOPS band to be brighter by $\sim$ 10 \% than the TESS photometric band, largely because CHEOPS is sensitive to optical wavelength scattering due to cloud particles at wavelengths 0.33-0.50 $\mu$m. 
We suggest that by comparing CHEOPS and TESS geometric albedos, qualitative information about the cloud particles of an exoplanet can be derived.
A similar suggestion was stated by \citet{Placek2016} which modelled disentangling scattered and emitted components from a combined Kepler and TESS photometry for revisited Kepler field planets.
If the CHEOPS band is brighter than the TESS bandpass, our modelling suggests that an optical wavelength scattering cloud particle is present.
While if the TESS band is brighter or nearer in brightness to the CHEOPS band, this is suggestive of a cloud-free or less/patchy cloudy planet since TESS is more sensitive to the near-IR thermal emission from an atmosphere than CHEOPS.
If different offsets in maximum flux in phase curves between the CHEOPS and TESS bands are present, this suggests a patchy cloud scenario, where CHEOPS is periodically more sensitive to optical wavelength scattering by clouds confined to one hemisphere of the planet compared to TESS.
We note that additional evidence (e.g. transit spectroscopy) is required to more conclusively infer cloud particle information.
However, this information may help identify planet candidates for more detailed spectrometer follow-up observations where observation time is more limited.

Our model generally supports the conclusions of \citet{Parmentier2016} which, in their GCM model grids, show that planets at the effective temperature of HD 189733b (T$_{\rm eq}$ $\sim$ 1200 K) would be bright in the Kepler band owing to the scattered light from cloud particles.
Our model agrees well with their non-cold trapped 0.1 $\mu$m particle Kepler band A$_{g}$ $\sim$ 0.20 scenario, but is less consistent with the 0.1 $\mu$m particle cold trap scenario, where A$_{g}$ $\sim$ 0.45. 
Our model is also reasonably consistent with their 1.0 $\mu$m particle size scenarios, where A$_{g}$ $\sim$ 0.25 $-$ 0.30.
We are also able to reproduce their prediction of a zero/small westward offset in the Kepler band for the T$_{\rm eq}$ = 1200 K model for all particle size and cold trap scenarios.
However, they suggest that MnS[s] and Na$_{2}$S[s] cloud materials may be more dominant than a silicate cloud (modelled here) for HD 189733b, possibly leading to different scattered light behaviour and variations in geometric albedo than those presented here.
Na$_{2}$S[s] condensation may also alter the gas phase Na abundance between the planet limbs and the dayside (where Na$_{2}$S[s] is likely to evaporate), which would affect the albedo spectra.

\section{Summary and conclusions}
\label{sec:Conclusions}

In this investigation we have examined the scattering and emission properties of our HD 189733b 3D cloud forming RHD simulation from Paper I.
We developed a Monte Carlo radiative transfer forward model for post-processing of atmospheric GCM/RHD output to produce combined scattered and emitted light observables.
We have shown that the wavelength-dependent 3D inhomogeneous cloud opacity and scattering behaviour of cloud particles can produce differences in phase curve behaviour between the east and west hemispheres of the planet.
Our combined scattered and emitted light simulations are consistent with available HST and Spitzer observations, with deviations most likely occurring because of to a decreased H$_{2}$O abundance from solar \citep[e.g.][]{Crouzet2014}, an unknown extra absorber at $\sim$ 0.5 $\mu$m \citep{Evans2013, Barstow2014}.
We also predict that the CHEOPS bandpass will be $\sim$ 10\% brighter than the TESS bandpass, because of the sensitivity of CHEOPS to optical wavelength scattering compared to TESS.
We suggest that qualitative information about the nature of the cloud particles on exoplanets can be inferred by comparing the relative geometric albedos and phase curve offsets of TESS and CHEOPS photometry data.
However, we emphasise the importance of obtaining high quality spectroscopic data in quantitatively inferring the cloud properties and atmospheric composition of exoplanet atmospheres.
Specific conclusions of our HD 189733b model results include

\begin{itemize}
\item A strong Rayleigh and/or backscattering Mie scattering component is likely required to increase the geometric albedo to observed levels in the HST B Band STIS measurements
\item We find that the eastern hemisphere of our HD 189733b model is more dominated by single scattering events, resulting in a Rayleigh-like classical phase function, while the western hemisphere is likely to have multiple-scattering events, resulting in a more isotropic classical phase function
\item Combined scattered and emitted light SEDs show that our model HD 189733b planet is likely to be at a similar brightness across 0.3-5$\mu$m on the dayside of the planet.
\item IR wavelengths can show westward offsets in scattered light phase curves. However, in this case, the overall magnitude of light scattered at these wavelengths is low.
\item We suggest HD 189733b to have apparent geometric albedos A$_{g}$ $\sim$ 0.222, 0.205, and 0.229 in the Kepler, TESS, and CHEOPS photometric bands, respectively.
\item We suggest HD 189733b may exhibit a zero or small ($\sim$ -5\degr) westward offset in the TESS and CHEOPS photometric bands.
\end{itemize}

The Monte Carlo radiative-transfer model framework presented here can be further extended to model other observable properties, such as transmission spectra, and other photon microphysical processes, such as photochemical haze formation.
Three-dimensional observables of 1D atmospheric models may also be readily simulated using this MCRT framework.

\begin{acknowledgements}
We thank the anonymous referee for constructive and insightful suggestions, which improved the manuscript and MCRT model substantially.
GL and ChH highlight the financial support of the European community under the FP7 ERC starting grant 257431.
Our local HPC computational support at Abu Dhabi and St Andrews is highly acknowledged.
We thank A. Collier-Cameron and D. Futyan for help with CHEOPS bandpass data.
We thank A. Mortier, K. Hay, and current/former members of the LEAP team for insightful discussions and suggestions.
Most plots were produced using the community open-source Python packages Matplotlib, SciPy, and AstroPy.
\end{acknowledgements}

\bibliographystyle{aa}
\bibliography{bib2}{}

\end{document}